\documentclass[12pt]{iopart}

\usepackage{iopams} 

\usepackage{amsmath}
\usepackage{graphicx} 
\usepackage{hyperref}

\begin{document}

\title{Emergent properties and the multiscale characterization challenge in condensed matter, from crystals to complex materials: a Review}

\author{Elisabetta Nocerino$^1$$^2$}

\address{$^1$ Department of Chemistry, Stockholm University, SE-10691 Stockholm, Sweden}
\address{$^2$ PSI Center for Neutron and Muon Sciences, 5232 Villigen PSI, Switzerland}

\ead{elisabetta.nocerino@su.se}
\vspace{10pt}
\begin{indented}
\item[]
\end{indented}

\begin{abstract}



The complexity of condensed matter arises from emergent behaviors that cannot be understood by analyzing individual constituents in isolation. While traditional condensed-matter approaches—developed primarily for ideal crystalline solids—have provided deep insights into symmetry, order, and electronic structure, they fall short in describing the rich, multiscale organization of hierarchical and soft materials. These systems exhibit structural correlations across multiple length and time scales, often governed by nonlinear interactions that span from molecular to macroscopic domains.

This review explores how the convergence of emerging experimental and computational strategies are redefining our ability to characterize and model such systems. We examine how multimodal techniques—combining scattering, imaging, and spectroscopy—can map structural order and dynamics across scales, with methods like small-angle scattering tensor tomography, dark-field imaging, and ultrafast spectroscopies providing unprecedented spatiotemporal resolution. On the computational front, machine learning approaches such as graph neural networks, neural operators, and physics-informed models offer powerful tools to connect disparate scales and uncover hidden correlations in high-dimensional data.

These advancements have the potential to close the gap between structure and function in complex materials, thereby addressing one of the grand challenges of contemporary material science: understanding and engineering multiscale architectures, whose emergent properties underpin the behavior of next-generation functional materials, biological systems, and adaptive technologies.

\end{abstract}

%
%
%
%
%

\section{Introduction}

The structural organization of matter is a fundamental theme that underlies many major research challenges across vastly different fields of modern science. This is because, although the behavior of the elementary building blocks of ordinary matter is quite well described within the current paradigms of particle physics, when they come together in larger structures new kinds of descriptions become necessary. A single atom cannot manifest superconductivity, a single segment of a polymer cannot exhibit viscoelasticity, a single neuron cannot produce consciousness. Central to this understanding is the concept of emergence, in which novel behaviors and functions arise from collective interactions between structurally organized individual components. In many-body systems, these emergent properties are not merely the sum of their parts and not necessarily can be understood or derived from knowledge of individual constituents alone, contrary to the claims of traditional reductionism \cite{feynman2017character,dirac1929quantum}.
This shift in perspective has been foundational in the field of Condensed Matter Physics \cite{anderson1972more}, which was initially focused almost exclusively on crystalline solids. During the early 20th century, significant progress was made in studying these kinds of systems by applying to atomic lattices the, at that time newly formulated, quantum mechanics approach. However, quantum mechanics alone was not sufficient as understanding phenomena like, e.g., magnetism, also requires statistical physics. This approach calculates the probabilities of a system's microscopic states at a given temperature, with collective properties emerging as averages. Small changes in thermodynamic parameters can lead to abrupt changes in these properties, known as phase transitions \cite{peliti2011statistical,stanley1971phase}. The study of such complex phenomena in "pure" and simple systems like crystals, which possess well-defined, periodic ordered structures, is what is considered the traditional focus of condensed matter physics. This focus often overlooked the inherent complexity in the “messier” forms of matter studied within the field of soft matter—broadly speaking, systems that are easily deformed by thermal fluctuations and external forces, characterized by structural features at mesoscopic scales—which cannot be fully understood using the singular approaches of conventional solid-state physics and often require case-specific, multidisciplinary investigation. 

These materials are not standardizable and exhibit a different kind of complexity that does not necessarily concern intricate quantum-level phenomena but rather the organization of the structures themselves, which, in some cases, is governed by physical principles of scaling and universality similar to those that drive phase transitions in systems like magnets \cite{de1979scaling}. Therefore, this complexity does not represent a lack of sophistication, but rather an opportunity to broaden our general understanding of the organization of matter in nature by including systems where order, functionality, and emergent properties arise from vastly different structural principles. After all, it is self-evident that matter, both living and non-living, generally self-organizes according to some kind of order, with the presence or absence of translational symmetry being a key discriminant factor. 
Translational symmetry refers to the property of a system whose structure is invariant under certain spatial translations \cite{nocerino2022comprehensive}. This is characteristic of materials with crystalline order, where atoms, molecules, or assemblies of molecules are arranged in a repeating lattice \cite{kittel2005int,ward2004structure}. Translational symmetry simplifies our investigation of the physics of these kinds of materials because it allows us to apply mathematical tools, like Bloch's theorem and Fourier analysis which, by exploiting spatial periodicity, reduce complex problems involving infinite many-body systems to more manageable calculations \cite{ashcroft1978solid}. 

Systems lacking translational symmetry are often labeled as "disordered". This classification is convenient because it allows us to circumvent the need to account for the underlying organizational principles of systems other than crystals. However, the absence of periodic repetition in space does not imply the absence of order and many such systems exhibit emergent behaviors arising from local interactions between their constituent elements. This raises intriguing questions about the conditions under which complexity emerges and whether there exists a critical point where increasing the number or strength of interaction between elements leads to spontaneous order within intricate structures \cite{bak1987self}.

The study of this kind of phase transitions involves non-equilibrium statistical mechanics and nonlinear dynamics, where spatially extended systems pose challenges well beyond those of low-dimensional systems. Here, concepts such as routes to chaos, bifurcations, and fractal dimensions describe how local interactions can generate order without a central organizing mechanism \cite{cross1993pattern}. This phenomenon is evident in nature where, e.g., network-based structures such as plant roots, neural networks, blood vessels, proteins or lightning patterns are frequently adopted to manage complex processes efficiently. This is done by exploiting hierarchical organizational features for which distinct, semi-independent functional units operating at different length-scales, interact with each other in a nested structure. These organizational strategies are adopted to optimize functionality and robustness of the system while minimizing usage of resources \cite{csermely2013structure}. 

The emergence of networks, as well as any other hierarchical organizational strategy, is due to the fact that Nature seems to favor decentralized systems, where global order arises from simple local rules, because they are able to adapt to environmental conditions, with properties often emerging not from design but from necessity.

Understanding these structural principles has recently become the focus of intensive research efforts in materials science, particularly in the fields of nanotechnology and polymer science \cite{whitesides2002self}. Here, we investigate how nanoscale and microscale interactions can produce macroscopic properties such as enhanced mechanical strength, self-healing capabilities, tunable transport properties, and energy-efficient processes. This understanding is intended to aid the design of self-assembling materials, metamaterials, and bioinspired systems that can mimic the ability of nature to self-organize, adapt, and maintain resilience under dynamic constraints\cite{zhang2016nano,fan2018creating}.

\begin{figure}[ht!]
  \begin{center}
    \includegraphics[width=0.95\linewidth]{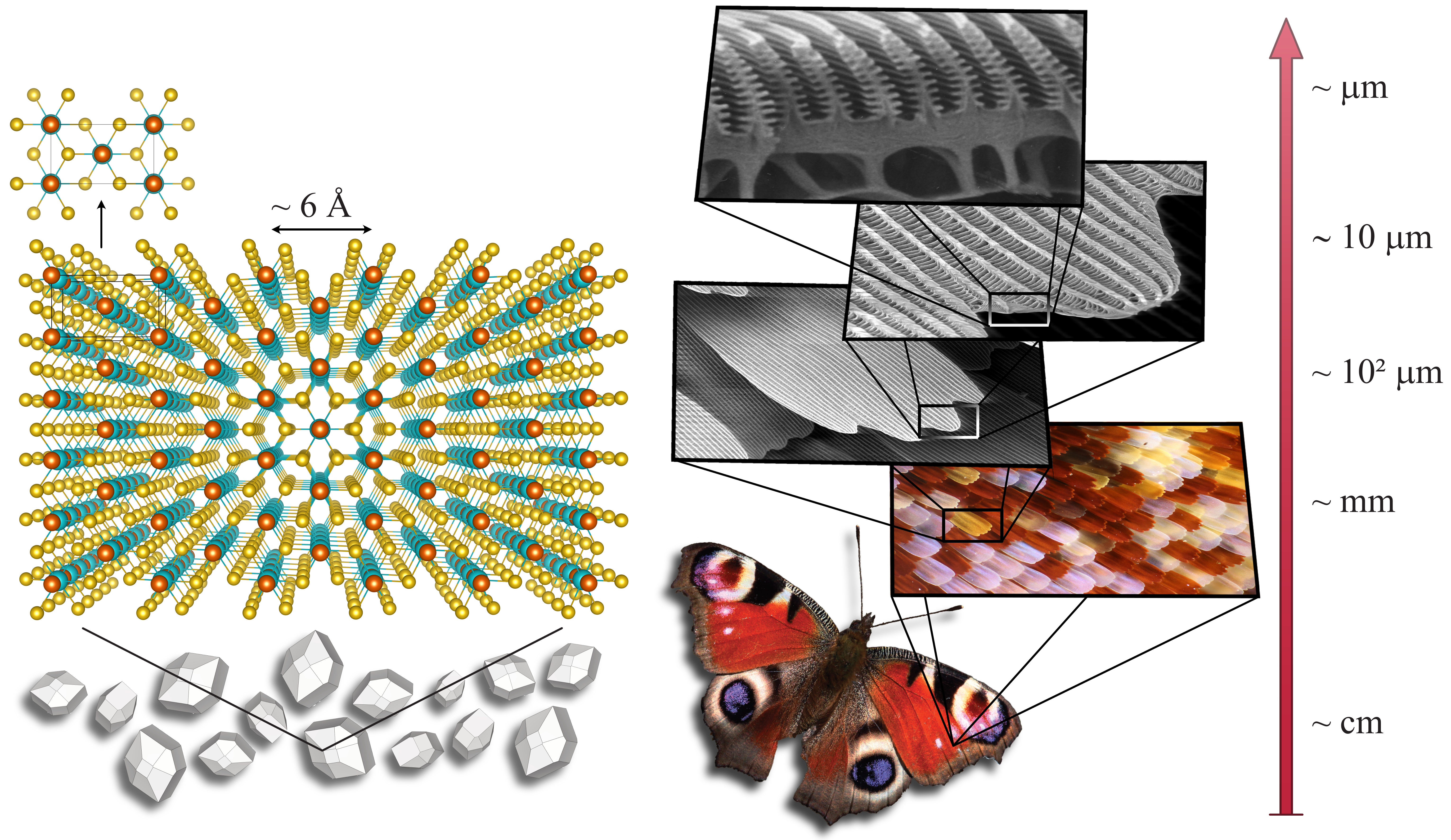}
  \end{center}
  \caption{Comparison between a system with translational symmetry and a system with hierarchical organization. (a) LiCrSe$_2$ crystal structure refined from diffraction data \cite{nocerino2023competition}, illustrating a system with a single level of organization. The unit cell repeats periodically in all directions, meaning the same structural motif defines the material at all scales. (b) Peacock butterfly wing, an example of hierarchical design, where structural features span multiple length scales. From the macroscopic specimen down to the nanoscale, successive magnifications reveal the complex arrangement of chitin-based micro- and nanostructures, imaged using scanning electron microscopy (SEM)\cite{butterfly1,butterfly2,butterfly3,butterfly4,butterfly5}. This hierarchical organization is responsible for the butterfly's structural coloration, lightweight architecture and mechanical strength.}
  \label{miller}
\end{figure}

While methods developed for the investigation of structure-property relationships in crystals traditionally provided valuable insights into material properties, applying them to materials with hierarchical organization, multiscale interactions, or non-periodic structures tends to be insufficient or requires significant adaptation. Moreover, while reliable structural data can be often provided with a single experimental method for crystalline materials (i.e., diffraction), complex materials require the integration of a wide variety of complementary characterization methods, together with new theoretical frameworks, that can account for the diverse organizational principles observed in natural systems (see Fig. \ref{miller}) \cite{strogatz2018nonlinear}.

On the computational front, traditional methods have so far driven theoretical advancements in physics. However, they mostly rely on simplifying assumption, such as giving periodic boundary conditions to their systems (effectively approximating any kind of material as a crystalline structure), to make calculations tractable \cite{kohn1965self}. This is because simulating every single particle and interaction in matter is computationally prohibitive due to the "curse of dimensionality", where the computational resources required grow exponentially with system size \cite{bellman1966dynamic}.

Recent developments in computational science, particularly in machine learning and artificial neural networks, allow to model complex non-linear relationships without explicitly simulating every single particle, and have been successfully applied to predict material properties \cite{chew2024advancing}, simulate molecular dynamics \cite{tamur2023artificial}, and analyze large datasets \cite{behler2007generalized}. 

On the experimental front, advancements in characterization techniques are driven forward by the need to address pressing societal challenges concerning environmental sustainability, technological development, and global health with novel functional materials \cite{osman2024synthesis}. Scattering methods, such as small-angle neutron and X-ray scattering (SANS/SAXS), have proven invaluable for investigating mesoscale structures \cite{narayanan2017recent} and, when combined with spectroscopic tools, they provide complementary insights into molecular composition, dynamics, and local environments. However, both scattering and spectroscopic techniques are inherently limited to probing a specific length scale in a single measurement and are not able to capture features and interactions that span multiple scales simultaneously. This poses non-negligible limitations when studying complex, hierarchical materials, where properties often emerge from interactions across nano-, meso-, and macroscopic domains. To address this characterization challenge, increasingly sophisticated imaging techniques are being developed to overcome the limitations of well established methods and enable a more comprehensive understanding of multiscale structures and dynamics \cite{busi2023multi,pfeiffer2008hard,liebi2015nanostructure,strobl2008neutron,peddie2022volume,huang2021three}.

The convergence of experimental and computational advancements creates an opportunity to link hierarchical material structures with their functional properties, by providing insights that can guide the design of materials that align with the current societal demands for sustainability and functionality.

In this paper, we provide an overview of structural order and emergent properties, focusing on how differently these concepts manifest themselves and can be characterized in systems with translational symmetry and hierarchical structures. We then discuss the limitations of traditional approaches and explore how these challenges are being addressed through emerging experimental and computational frameworks that incorporate multiscale interactions and hierarchical organization.

In order to address the complexities of emergent properties and multiscale characterization in the contemporary landscape of materials science, it is necessary to depart from traditional, compartmentalized approaches and embrace a structuralist perspective that allows to see the interrelations and overarching frameworks that define various systems. This perspective invites us to zoom out, to see how different systems echo each other in structure and function, even when they arise in completely different contexts. More often than not, the most interesting solutions come from unexpected places and a method from one field can very well end up being the key to a problem in another. This review reflects that belief, bringing together concepts from crystallography, soft and hierarchical matter, statistical physics, computational modeling, and data-driven approaches, in the hope of uncovering solutions that might remain hidden within the boundaries of a single discipline. This review aspires to serve as a conduit for such interdisciplinary dialogue, offering a synthesis of historical context, foundational concepts, and the latest experimental and computational advancements. 

Through this integrative approach, the author hopes to equip readers with some of the knowledge and tools necessary to navigate and contribute meaningfully to the rapidly evolving and multidisciplinary field of materials science. It should be mentioned that the distinctions discussed in this review are informed by perspectives gained by the author herself through transitions between traditionally separate research domains, particularly from solid-state physics to soft matter, where the criteria for structural order and complexity often differ substantially.


\section{Structural order}
\label{order}

In materials science, the concept of structural order is a foundational principle in both solid-state and soft matter physics, yet its interpretation diverges significantly between these fields due to the inherent differences in the systems they study. In solid-state physics, order is rigorously defined by the precise and periodic arrangement of atoms in a crystalline lattice, where any minor deviation such as a missing atom or a defect is regarded as "disorder". This strict definition stems from the fact that symmetry and periodicity affect electronic properties in crystals by creating a periodic potential that organizes electron energies into bands and gaps. Symmetry governs the degeneracy and anisotropy of these bands and consequently determines how quantum states are allowed to interact under symmetry-preserving conditions \cite{grosso2013solid,kittel2018introduction,ashcroft1978solid}. This enables the emergence of quantum phenomena such as magnetism, where symmetry breaking between spin states can lead to ordered magnetic phases \cite{tinkham2003group}, or superconductivity, where the symmetry of the electronic band structure determines the pairing mechanisms and the nature of the superconducting gap \cite{higgs1966spontaneous,nambu1995broken}, as well as topological states, where symmetry-protected features in the band structure, such as Dirac or Weyl points, give rise to robust surface states and unconventional electronic excitations \cite{senthil2015symmetry}. These kinds of phenomena are traditionally in the focus of solid state condensed matter physics.

Soft matter systems tend to find their order at intermediate scales, in mesoscopic organization rather than in atomic-scale periodicity. Here, the notion of structural order is considerably broader and more flexible compared to its counterpart in crystalline solids. For example, the alignment of polymer chains, the orientational organization of liquid crystals, or the arrangement of colloidal particles may be considered "ordered," even if the overall organization is far from perfect and the positions of individual components (let alone single atoms) are irregular or fluctuating \cite{jones2002soft}. This is because, in soft matter systems, "order" acquires meaning only in connection to its functional implications. These forms of order are statistical, driven by local interactions and thermal fluctuations within components that collectively create patterns in orientation, density, or connectivity that eventually produce macroscopic properties \cite{de1979scaling}. 

As will be discussed in section \ref{emergent}, this dynamic and emergent nature of order allows for phenomena such as phase transitions in liquid crystals, driven by changes in molecular alignment under external fields \cite{singh2000phase}, or elasticity, where the network-like arrangements of polymer chains respond to mechanical forces \cite{hart2021material}. These behaviors are often tied to the interplay between order and disorder, as the systems remain functional and robust even under far-from-equilibrium conditions. The focus of soft matter physics lies in understanding these emergent patterns and their implications for both fundamental processes and practical applications \cite{jones2002soft}. 

In this section, we explore the principles of symmetry and periodicity in crystals and extend the discussion to the different kinds of structural order found in soft matter systems.

\subsection{Periodicity, Symmetry Representations and Reciprocal Lattice in Crystals}

In three-dimensional space, symmetry operations are transformations whose action leaves an object unchanged. These operations are associated with symmetry elements, which include: rotation axes (an object has an n-fold rotation axis if it can be rotated by 360$^\circ$/n and remains indistinguishable from its original orientation); mirror planes (reflection across a plane where the object appears unchanged); inversion centers (points in space such that every part of the object has an equivalent part directly opposite and equidistant from the inversion center).
Combinations of these symmetry operations can produce more complex transformation like roto-inversions and roto-reflections, and the complete set of symmetry operations that leave at least one point unmoved defines the object's point group \cite{tinkham2003group}.

To describe the organization of matter in crystalline solids, however, we must consider translational symmetry as well. This involves shifting the entire structure by specific vectors so that it overlaps perfectly with its original position. In a crystalline solid, atoms are arranged in a repeating pattern throughout space, forming a crystal lattice. The smallest repeating unit in this lattice is the unit cell, defined by vectors \textbf{a}, \textbf{b}, \textbf{c} and angles $\alpha$, $\beta$, $\gamma$ \cite{hammond2015basics} and the specific choice of unit cell is guided by the aforementioned symmetries of the crystal. Bravais lattices are the fourteen unique three-dimensional lattices that can be created using these symmetry operations combined with translational symmetries. They are classified into seven crystal systems based on the lengths and angles of the unit cell vectors: triclinic, monoclinic, orthorhombic, tetragonal, trigonal (rhombohedral), hexagonal, and cubic \cite{wondratschek2004international}.

When point symmetries are combined with translational symmetries, we obtain space groups, which describe the full symmetry of a crystalline material. There are only 230 unique space groups, each representing a distinct combination of symmetry operations, which fully describe the arrangement of the periodic tridimensional patterns that can occur in any possible crystal \cite{wondratschek2004international}.

This periodicity lays the foundation for the concept of reciprocal lattice, a mathematical abstraction that greatly simplifies the investigation of crystalline systems. It is constructed as the Fourier transform of the direct lattice, having reciprocal lattice vectors $\boldsymbol{a^*}$, $\boldsymbol{b^*}$, $\boldsymbol{c^*}$, normalized by the scalar triple product of the direct lattice vectors. These vectors are orthogonal to the plane defined by the cross product of the remaining pair of direct lattice vectors (i.e., $\boldsymbol{a^*} \propto \boldsymbol{b} \times \boldsymbol{c}$). The reason for the existence of this kind of mathematical construct is that, in order to probe the organization of matter in materials, we rely on scattering methods. Specifically, in the case of crystalline systems, we use diffraction, for which probe particles such as neutrons, X-rays or electrons, scatter elastically with the atomic centers in the material and are consequently deflected under a limited number of well-defined angles \cite{nocerino2022comprehensive}. Here, because of the periodic organization of atoms in well defined crystalline planes (i.e., the Miller planes), the probing radiation will be collectively reflected and create interference patterns, similar to light through a set of slits, by virtue of their wave-particle duality. Reflections through sets of mutually parallel Miller planes will give rise to constructive interference resulting in a diffraction spot (i.e., an intensity peak) whose intensity and angular position is directly connected to the specific set of planes via the Bragg's law:

\begin{eqnarray}
 n\lambda = 2d_{hkl}\;\; \sin\theta.
\label{bragg}
\end{eqnarray}

Here, $\lambda$ is the wavelength of the incident radiation; $d_{hkl}$ is the spacing between parallel Miller planes where the indices ($h$,$k$,$l$) uniquely identify the family of planes of equation $m = h\boldsymbol{a} + k\boldsymbol{b} + l\boldsymbol{c}$, which intercept the crystallographic axes in the points $h\boldsymbol{a}$, $k\boldsymbol{b}$, $l\boldsymbol{c}$; $\theta$ is half of the angle formed by the incident and diffracted radiation with the family of scattering planes. Therefore, diffraction peaks, as they result from the constructive interference of scattered waves, can be understood as the Fourier transform of the periodic electron density (or scattering potential) in the real space crystal. From the positions and intensities of these peaks, one can determine lattice parameters, symmetry, and atomic form factors that ultimately allow to deduce how atoms are positioned in the unit cell \cite{nocerino2023multiple,nocerino2022nuclear}. Given that diffraction patterns are basically what we measure when we investigate the structure of crystalline systems, and since the reciprocal lattice is the Fourier transform of the direct lattice, it is easy to see that reciprocal space provides a natural domain to understand the periodic organization of matter in crystals. Moreover, as mentioned above, reciprocal lattice vectors are obtained as vector products of real lattice vectors. Here, it is possible to define a vector $\boldsymbol{d*_{hkl}}$ reciprocal of the inter-planar distance in the direct space $d_{hkl}$, which will have direction orthogonal to the family of planes identified by the triad ($h$,$k$,$l$): 

\begin{eqnarray}
\boldsymbol{d^*_{hkl}} = h\boldsymbol{a^*} + k\boldsymbol{b^*} + l\boldsymbol{c^*}.
\label{}
\end{eqnarray}

The reciprocal lattice vector $\boldsymbol{d^*_{hkl}}$, given as a linear combination of the vectors $\boldsymbol{a^*}$, $\boldsymbol{b^*}$, $\boldsymbol{c^*}$, connects points in reciprocal space just like $d_{hkl}$ connects planes in the real space. This aspect clarifies why the formalism of the reciprocal space is convenient: instead of dealing with entire families of planes, the reciprocal lattice condenses this information into discrete points of coordinates ($h$, $k$, $l$). This discretization of the reciprocal space leads to the definition of the Brillouin zone, the region in reciprocal space that contains all the possible wave vectors required to describe the periodic properties of a crystal. Every reciprocal lattice point constitutes the center (and origin) of a Brillouin zone, with high-intensity elastic Bragg peaks occurring at these central points. The Brillouin zone is constructed by connecting the origin of the zone to the nearest neighboring reciprocal lattice points using the shortest possible translation vectors. The region of space enclosed by the polyhedron formed by intersecting these vectors at their midpoints with perpendicular planes defines the First Brillouin zone of the crystal \cite{ashcroft1978solid}. This geometric definition allows to exploit the translation symmetry of crystalline system, because every wave vector outside the Brillouin zone can be represented within the zone by translation with a reciprocal lattice vector, and reflects all the symmetries of a given lattice in reciprocal space. 

The power of this approach is clear if we consider that the periodicity of crystalline systems restrict any kind of calculation to the fundamental repeating unit (i.e., the unit cell) and symmetry operations can be mathematically represented using matrices that transform the coordinates of points within the crystal lattice. These representations simplify the understanding of the effects of symmetry operations on crystal structures (and their physical properties, e.g., magnetic ordering \cite{nocerino2023unusually,matsubara2020neutron}), that can be retrieved in all possible cases from straightforward linear algebra operations. Linear algebra operations are particularly efficient in these systems because symmetry reduces the number of independent variables and allow properties arising from the specific electronic band structures to be computed within the Brillouin zone alone \cite{kratzer2019basics}. This approach is so convenient that it is also applied in soft matter fields when possible. For instance, proteins, that are complex macromolecules not inherently crystalline in their native biological states, can be arranged into highly ordered lattices through crystallization processes \cite{durbin1996protein}. The availability of protein crystals facilitates the determination of their three-dimensional structures via standard diffraction. Moreover, using symmetry-adapted basis sets simplifies calculations by restricting them to only the interactions allowed by the crystal's symmetry \cite{rodriguez2001magnetic}. This drastically reduces the number of variables and eliminates redundant computations, thereby increasing dramatically the computational efficiency in applications like density functional theory and phonon dispersion analysis.

\subsection{Types of Structural Order in Soft Matter Systems}
Soft condensed matter lacks a universally agreed-upon definition, but one possible description is this: a material qualifies as soft condensed matter if it exhibits strong resistance to compression and weak resistance to shear forces, and can then be easily deformed \cite{cates2004soft}. Structural order in this context is a broad concept that refers to the degree of spatial organization or regularity in the arrangement of molecular or mesoscopic components, and varies significantly depending on the specific system. Here, the concept of "degree of crystallinity" is typically used to indicate the portion of the material that adopts a regular, repeating arrangement. Unlike crystalline solids, which have strict long-range order and high degrees of crystallinity (ideally 100$\%$), soft matter systems exhibit varying degrees of crystallinity that coexist with amorphous or disordered regions \cite{uzun2023methods}. These systems can also display varying degrees of short-range or intermediate-range order, which arise from their molecular or mesoscopic structures and are influenced by factors such as flexibility, thermal motion, and weak interactions. Short-range order refers to local regularity, such as consistent bond lengths and angles over a short distance that does not extend to the entire material \cite{cowley1950approximate}. Intermediate-range order is typical for materials like glasses and gels that exhibit structural patterns extending beyond the nearest neighbors, but do not repeat periodically \cite{singh2023intermediate}. It should be noted that while glasses are mechanically rigid, their lack of periodicity distinguishes them from crystalline solids, and their formation involves slow relaxation dynamics. This conceptual similarity to soft matter systems is the reason for their inclusion in this discussion.

Based on the degree and type of structural order, we group soft condensed matter into three main categories: polymers, which exhibit primarily short-range order influenced by stereoregularity; liquid crystals, which possess intermediate-range order with varying degrees of orientational and positional alignment; and gels/glasses, where short- to intermediate-range order arises from networked or locally coordinated structures without long-range periodicity. Transversal to these foundational categories is hierarchical order, which underlies the multiscale structural organization of many soft materials.

\subsubsection{Polymeric structures}
Polymers are long chain-like molecules made up of repeating monomer units. The type of structural order associated with this kind of structure is stereoregularity, which refers to the systematic spatial arrangement of side groups or atoms along the backbone of the polymer chain. Stereoregularity determines the polymer's ability to crystallize as well as its physical properties \cite{szweda2023sequence}, in fact, polymers can be categorized based on stereoregularity \cite{natta1959properties}. 

In isotactic polymers, all substituent groups are oriented on the same side of the polymer backbone. This uniform arrangement causes a close packing of the polymer chains, leading to higher degrees of crystallinity, with consequently higher melting points and mechanical strength.

In syndiotactic polymers, substituents alternate regularly between opposite sides of the polymer backbone. This configuration also allows for significant crystallinity, though the resulting properties may differ from those of isotactic polymers due to differences in packing efficiency. Syndiotactic polystyrene, for instance, displays high crystallinity and thermal stability, which make it widely utilized for high-temperature applications \cite{pasztor1991thermal}.

Atactic polymers have randomly oriented substituents, resulting in amorphous structures with little to no crystallinity. This structural arrangement allows for high flexibility and lower melting points, but may lack mechanical strength.

\subsubsection{Liquid Crystals}

Liquid crystal identifies a distinct state of matter that exhibit structural order intermediate between that of isotropic liquids and crystalline solids. This peculiar phase behavior arises from the anisotropic shape of the liquid crystals molecules, which align in specific patterns while retaining some fluidity (as in, e.g., nematic liquids \cite{schutz2015rod}. Liquid crystalline structures can be described by three main parameters that quantify different types of order \cite{stephen1974physics}: orientational order, which measures the degree to which molecules align along a common direction (the director) over long ranges; positional order, which indicates the extent of translational symmetry in the molecular arrangement, similar to crystalline materials; and bond-orientational order, which, though formalized later than the other two \cite{nelson1979dislocation}, describes the alignment of the bonds (lines connecting neighboring molecular centers) rather than the positions of the molecules themselves. This reflects long-range order along these lines without requiring regular spacing, resulting in short-range positional order along the line itself.
Liquid crystal compounds often exhibit polymorphism, meaning that they can exist in multiple phases within the liquid crystalline state, known as mesophases \cite{de1993physics}. These mesophases differ in their degrees of order, which may be restricted to one or two dimensions or allow some translational motion.

In the nematic phase, molecules align along a common axis, while lacking positional order. This phase is widely used in liquid crystal displays (LCDs) due to its optical anisotropy and fast response to electric fields \cite{schadt1971voltage}.

In the smectic phase, liquid crystals form layered structures, with partial translational invariance, having the molecules in each layer maintaining orientational order. Different smectic phases (typically labelled with latin letters, similar to crystallographic space groups), are distinguished by the specific arrangement and orientation of molecules both within the layers and relative to the layers. Smectic A and Smectic C for example are common subphases, with the latter exhibiting tilted molecular arrangements within the layers \cite{de1993physics}. These phases are studied for applications in advanced display and sensor technologies \cite{clark1980submicrosecond}.

In the cholesteric phase, molecules of the liquid crystal exhibit orientational order similar to the nematic phase but also exhibit a chiral twist, leading to a helical structure. This helical structure has a well-defined periodicity, known as the pitch, which gives rise to unique optical properties, such as selective reflection of light at specific wavelengths. This property is exploited in thermochromic materials and photonic devices \cite{belyakov1979optics}.

\subsubsection{Gels and Glasses}
Gels and glasses lack long-range periodicity but exhibit structural order at smaller scales \cite{cowley1950approximate}. In gels, the order often arises from a network of interconnected molecules or particles that form a three-dimensional structure, which immobilizes and traps solvent molecules within its matrix. This network may be formed through chemical bonds, such as covalent crosslinks in polymer gels, or through physical interactions due to the electromagnetic interactions between the molecular structures involved in physical gels \cite{de1979scaling}. The trapped solvent molecules contribute to the gel’s high liquid content and enable its viscoelastic properties, resulting in a material that exhibits both solid-like stability and liquid-like deformability. In glasses, short- and intermediate-range order is characterized by bond angles and distances that closely resemble those found in their crystalline counterparts, reflecting local structural motifs or coordination environments typical of the chemistry of the specific material. These local arrangements result from the same fundamental bonding principles that govern crystalline structures, such as covalent or ionic interactions, and can include features like tetrahedral coordination in silica-based glasses or octahedral units in transition-metal oxide glasses. However, unlike crystals, where these motifs repeat periodically over long distances, glasses lack translational symmetry, and the organization beyond a few atomic or molecular lengths becomes random. This peculiar structure is responsible for the optical transparency typical of glasses, as the random arrangement of atoms prevents scattering of visible light \cite{anderson1995through}.

\subsection{Self-Assembled Structures and Hierarchical order}\label{self-assembled}
Self-assembly is a hallmark of soft matter systems and refer to organized arrangements of molecules, particles, or other components that form spontaneously due to the intrinsic interactions among the system's constituents, without the need for external manipulation \cite{whitesides2002self}. These structures are governed by the minimization of the system's free energy and are typically driven by a balance of weak forces such as van der Waals interactions, hydrogen bonding, electrostatics, hydrophobic interactions, and steric effects. In fact, many self-assembled structures are dynamic and able to disassemble or reassemble in response to environmental stimuli, repair damage, and adapt to changing conditions (e.g., temperature). Natural self-assembly is inherently hierarchical, meaning that the structural organization process occurs in a stepwise fashion across multiple length scales, from molecular interactions to macroscopic structures \cite{scacchi2021self}.  This multiscale organization is a defining characteristic of biological systems and allows nature to create sophisticated structures with diverse mechanical, optical, and chemical properties (see Fig. \ref{miller}). At the same time, it minimizes the energy and resources needed to build them, by relying on simple molecular building blocks.

At the molecular level (lengthscale range $\sim$[1 Å - 1 nm]), self-assembly begins with fundamental interactions that organize molecules into primary structures. To mention a few notable examples, biopolymeric cellulose chains are formed through covalent bonds between glucose units, stabilized by hydrogen bonding \cite{frey1954fine}. DNA organizes into its double-helix structure via base-pairing and $\pi$-$\pi$ stacking interactions \cite{watson1953molecular}. Proteins fold into secondary structures such as $\alpha$-helices and $\beta$-sheets through hydrogen bonding along the polypeptide backbone \cite{pauling1951structure}. Collagen, like other proteins, begins as polypeptide chains composed of amino acids arranged in a characteristic repeating sequence, Gly-X-Y, where X is often proline and Y is often hydroxyproline \cite{ramachandran1954structure}. Lipid molecules self-assemble into bilayers through hydrophobic and hydrophilic interactions \cite{singer1972fluid}. These molecular structures represent the foundational components for higher-order assemblies.

At the nanoscale level (lengthscale range $\sim$[1 nm - 10$^2$ nm]), molecular assemblies form larger structures through cooperative interactions. Cellulose chains aggregate into microfibrils, creating a nanoscale framework for plant cell walls. DNA associates with histone proteins to form nucleosomes, and compact the molecule into a higher-order structure. Proteins assemble into tertiary and quaternary structures, such as enzyme complexes or hemoglobin. Another example is the self-assembly of virus capsids, composed of protein subunits which self-assemble into highly symmetric shells that protect the viral genome and minimize the genetic information needed for their formation \cite{caspar1962physical}. Collagen triple helices self-assemble, producing fibrillar structures. Lipid bilayers curve and close upon themselves to form vesicles or micelles.

At the supramolecular level (lengthscale range $\sim$[10$^2$ nm to 10 $\mu$m]), self-assembled units develop into interconnected or compartmentalized systems. Cellulose microfibrils aggregate into larger fibers, contributing to the mechanical stability of biological materials. DNA, compacted into nucleosomes, organizes further into chromatin fibers. Protein assemblies transition into larger complexes, such as those forming cytoskeletal networks in cells. Collagen fibrils aggregate into fibers, providing structural support to connective tissues. Vesicles organize into multi-vesicular structures or layered membranes.

At the mesoscopic level (lengthscale range $\sim$[10 $\mu$m - 1 mm]), hierarchical assemblies result in systems with defined mechanical or functional roles. Cellulose fibers integrate into macroscopic structures like plant stems or leaves. Chromatin fibers further condense into chromosomes. Proteins such as actin and tubulin self-assemble into cytoskeletal filaments that provide structural integrity and dynamic transport pathways in cells. Collagen fibers combine with other biological components to form connective tissue networks such as tendons, cartilage or bone. Vesicle networks organize into functional compartmentalized systems, including emulsions stabilized by proteins or lipids.

At the macroscopic scale [$\geq$ mm], fully developed self-assembled systems emerge with large-scale functionality. Cellulose fibers come together to ensure the structural integrity of entire plants. DNA, organized into chromosomes, is housed within cells that aggregate and specialize, forming the foundation for the complex structures and functions of living organisms. Protein assemblies organize into tissues or large-scale networks supporting biological functionality. Collagen-based structures combine with mineralized components such as hydroxyapatite crystals to form bone, achieving high strength and resilience for structural support. Vesicle-based systems form bulk materials such as mayonnaise or synthetic hydrogels. 

Hierarchical self-assembly in nature inspires the design of biomimetic materials such as artificial scaffolds for tissue engineering and structural components, which mimic the multiscale architecture of collagen or the mechanical strength of fibrous networks like plants, spider webs and insect exoskeletons \cite{rodrigues2016bioinspired}. Nanomaterials like photonic crystals are engineered to emulate the structural color seen in butterfly wings or peacock feathers \cite{zhao2012bio}, and hydrogels, that replicate the dynamic self-assembly of biological systems, are used in drug delivery and wound healing \cite{webber2017drug}.

\subsection{Reciprocal space in soft matter} \label{Reciprocal space in soft matter}
The concept of reciprocal space is defined for, and deeply rooted in, the periodic nature of crystalline solids, as discussed above: the Fourier transform of the atomic lattice yields a reciprocal lattice, where discrete points in reciprocal space correspond to well-defined periodicities in real space. Conceptually, this framework does not adapt well to the description of systems where long-range periodicity (and therefore translational symmetry) is absent. Historically, direct imaging techniques (like electron or optical microscopy) lacked the resolution to observe atomic or molecular arrangements and scattering was the only method that could provide indirect but precise insights into the structure of condensed matter.
Scattering experiments are still among the most powerful tools to investigate the properties of matter, and they naturally produce data in reciprocal space, as mentioned above. Therefore, this approach was extended to systems lacking translational symmetry exploiting the general applicability of the Fourier transform which, also in this case, can relate spatial correlations in real space to patterns in reciprocal space.
The spatial arrangement of particles or molecular units in soft matter can be described by the density function $\rho(\boldsymbol{r})$, whose Fourier transform defines the structure factor $S(\boldsymbol{q})$:

\begin{eqnarray}
S(\boldsymbol{q}) = 1 + \int [g(r) - 1] e^{-i \boldsymbol{q} \cdot \boldsymbol{r}} d\boldsymbol{r},
\label{strucfac}
\end{eqnarray}

where $g(r)$ is the pair correlation function, which describes the probability of finding a pair of particles separated by a distance $r$, normalized by the average density, and $\boldsymbol{q}$ is the scattering vector in reciprocal space, which is the inverse of a characteristic real-space distance $\boldsymbol{r}$ \cite{hansen2013theory}. This expression describes how correlations between particles (given by $g(r)$) manifest as features in reciprocal space via their specific scattering pattern.

Soft matter systems are commonly studied using scattering techniques with different probes (X-ray, neutron, or light), where the intensity of scattered waves is measured as a function of $\boldsymbol{q}$, which contains the spatial frequency information. In crystalline materials, long-range periodicity results in sharp Bragg peaks at discrete $\boldsymbol{q}$-values, corresponding to the well-defined lattice spacings. In soft matter systems without strict periodicity, the interference of scattered waves is no longer constructive at discrete points. Instead, the scattering intensity becomes a continuous function of the scattering vector $\boldsymbol{q}$, where broad features indicate characteristic length scales, such as particle-particle distances or domain sizes. By way of example, particles in colloids or suspensions tend to aggregate due to interparticle interactions, and form clusters that create characteristic length scales that manifest as broad peaks in the scattering profile \cite{kashanchi2023using}. Similarly in liquids, pair correlations between neighboring atoms create broad peaks corresponding to average interatomic distances \cite{loo2021uncovering}. In glasses, intermediate-range order generates weak peaks corresponding to structural motifs or local coordination environments \cite{meyer2004channel}. In systems that are completely disordered or dominated by thermal fluctuations, the scattering intensity is diffuse and does not show any specific feature, which indicates the absence of well-defined spatial correlations \cite{welberry2022diffuse}. 


\section{Emergent Phenomena and the Concept of Multiscale}
\label{emergent}

Emergent phenomena occur when a system’s properties cannot be reduced to the sum of its individual components. These properties arise from interactions among the components, giving rise to novel states that are independent of the specific details of the parts. This independence confers universality to emergent properties, enabling vastly different systems to exhibit similar behaviors \cite{anderson1972more}.

Emergent properties are considered irreducible because they cannot be fully explained by examining the system’s building blocks in isolation. This irreducibility makes emergent behaviors inherently difficult (if not impossible) to predict solely from knowledge of the individual components and their fundamental interactions. For example, the Bardeen–Cooper–Schrieffer (BCS) theory of superconductivity provided a detailed understanding of the phenomenon a posteriori, but the emergent property itself could not have been fully predicted a priori based solely on the properties of individual electrons \cite{bardeen1957theory}.

Given these characteristics, it is self-evident that emergent phenomena can only manifest at scales larger than those of individual components. These scales can be extrinsic, such as the number of parts, time scales, or spatial dimensions, or intrinsic, i.e., associated with the emergent property itself \cite{sep-properties-emergent}. Therefore, we could say that a scale is a way to describe a physical system with a specific characteristic dimension. In crystallography, there is typically only one characteristic extrinsic scale: the periodicity of the lattice, defined by the lattice translation vector, which reflects the geometric structure of the material. Here, new scales can emerge under certain conditions \cite{landau1937theory}. For example, in magnetic systems, microscopic agents such as individual atomic spins, initially described at the lattice scale, can align in response to thermodynamic stimuli, leading to the emergence of a macroscopic magnetic field. This process introduces a new periodicity, larger than the lattice (except in the case of ferromagnetic systems), represented by the propagation vector of the magnetic structure \cite{nocerino2023competition}. This new periodicity constitutes another extrinsic scale of the system, but the emergent magnetic field itself reflects an intrinsic scale associated with the coherence length of the magnetic order. This phase transition depends on thermodynamic conditions and occurs only below a critical temperature where breaking of rotational symmetry in the spin orientations leads to magnetic ordering \cite{blundell2001magnetism}. Above this temperature, the system reverts to a state described solely by the lattice scale.


In hierarchical structures, the emergence of multiple scales does not result from phase transitions in response to certain thermodynamic conditions but is present ab initio, because of the structural organization of these kinds of systems (see Fig. \ref{miller}b). These distinct characteristic scales interact non-linearly across hierarchical levels, even when the geometry of the structure itself is linear \cite{ahnert2017revealing}. Unlike crystalline systems, where the periodicity of the lattice imposes a linear structural behavior that can be modeled through harmonic approximations \cite{ashcroft1978solid}, hierarchical systems possess multiple interacting correlation lengths. 

Therefore, systems possessing translational symmetry are characterized by a single dominant structural scale, defined by a lattice-like periodicity, where emergent phenomena introduce new scales that remain closely tied to the lattice structure. Hierarchical materials, instead, are intrinsically multiscale and their emergent properties are shaped by the dynamic interplay and coupling across pre-existig scales. For this reason, when talking about emergent phenomena in condensed matter, it is always important to keep in mind what kind of system we are considering. Broadly speaking, we may classify emergent phenomena into three main categories depending on underlying interaction mechanisms and conditions under which emergence occurs \cite{halley2008classification}: emergence from phase transitions, non-equilibrium emergence, and intrinsic emergence in multiscale structures. 

\subsection{Emergence from phase transitions}
Phase transitions represent the classical framework of emergence \cite{landau1937theory}, marked by the appearance of new macroscopic states due to variations in thermodynamic parameters like temperature, pressure, or external fields. Here, the system evolves from an equilibrium state to another equilibrium state through a transition characterized by the competition between internal energy $U$ minimization (tendency to order) and entropy $S$ maximization (tendency to disorder), as the system seeks to minimize its free energy to reach the equilibrium condition \cite{jaeger1998ehrenfest}. Depending on the thermodynamic constraints, the appropriate definition of free energy will determine the equilibrium state of the system: the Helmholtz free energy $F = U - TS$ is used for systems at constant temperature $T$ and volume $V$, while the Gibbs free energy $G = U + PV - TS$ applies for systems at constant temperature and pressure $P$. For example, in temperature-driven phase transitions, at high temperatures, the entropy term dominates and disordered configurations are favored to minimize the free energy, while at lower temperatures the internal energy becomes more significant, driving the system toward a more ordered state. Both the internal energy and the free energy are state functions, meaning that they depend only on the initial and final states of the system, not on the path taken to reach these states. 
The paramagnetic-to-magnetically ordered transition, mentioned above, exemplifies this type of emergent phenomena. Such transitions involve rotational symmetry breaking, where the system goes from a more symmetric (but less ordered) state with random spin orientations, to a less symmetric (but more ordered) one with a preferred spin orientation direction, below a certain critical temperature \cite{blundell2001magnetism}. The degree of order in the system is quantified by an order parameter, which takes the value zero when the system is in the disordered state and becomes non-zero once the ordered state has emerged across the transition point (i.e., the critical point) \cite{landau1937theory}.
At the critical point, local fluctuations in physical properties (e.g., electron or spin density) become scale-invariant as the system loses any characteristic length scale near the transition. As a result, the system develops long-range correlations, where fluctuations at one point are correlated with those occurring far away. These correlations extend across the entire material and their spatial extent is quantified by the correlation length $\xi$. Near the critical point $\xi$ diverges to infinity, and the behavior of the system is governed by critical exponents, following scaling laws that describe how physical quantities depend on the distance from the critical point. For instance, in a magnetic system, one such quantity can be the magnetization $M(T)$ that, near the critical temperature $T_c$, follows a power law of the type $M(T) \propto (T_c - T)^{\beta}$. Here, $\beta$ is a critical exponent that allows to classify the specific phase transition under a certain universality class. These universal properties are powerful because they are independent of the microscopic details of the system and depend only on a few key factors, such as the dimensionality of the system and the nature of the interactions. This universality provides a unified framework to describe emergent phenomena across disciplines, because it allows systems with vastly different microscopic constituents to exhibit similar behavior near their respective critical points \cite{cheung2011phase}. To mention a few notable examples, the solidification of crystalline solids represents itself a classic phase transition, where atoms self-organize into periodic structures to minimize their potential energy, as the system overcomes the entropic preference for disorder that dominates in the liquid phase. Similarly, self-assembly in molecular systems provides another example of phase transition-driven emergence. Block copolymers, for instance, undergo microphase separation due to competition between enthalpic interactions, which favor segregation of different polymer blocks, and entropic contributions, which resist ordered configurations. This leads to the spontaneous formation of lamellar, cylindrical, or spherical domains that minimize the system’s free energy while introducing periodic patterns at the nanoscale \cite{jones2002soft}. Liquid crystals experience symmetry breaking in the same way as magnets do, with orientational order of their rod-like molecules emerging across the temperature dependent isotropic-to-nematic phase transition \cite{de1979scaling}. In binary alloys, compositional fluctuations become scale-invariant at the critical point and phase separation occurs, as the system minimizes its free energy by forming regions with distinct compositions \cite{cahn1958free}. In conventional superconductors, the superconducting transition is driven by the formation of bound electron pairs that collectively occupy a macroscopic quantum state (Cooper pairs). Within the framework of mean field theory, this transition is described by the BCS theory, according to which the system minimizes its free energy by establishing a pairing gap in the electronic density of states below a critical temperature $T_c$ \cite{bardeen1957theory}. The resulting symmetry breaking leads to the emergence of two fundamental length scales: the coherence length, which characterizes the spatial extent of Cooper pairs, and the penetration depth, which quantifies the expulsion of magnetic fields due to the Meissner effect.
At the transition point, the behavior of the superconducting order parameter follows universal scaling laws, similar to classical phase transitions, despite the underlying quantum nature of the system. 
A novel and unconventional use of the formalism of Landau's theory of phase transitions was recently used to estimate buckling critical pressures in collapsible tubes \cite{laudato2023buckling}.

\subsection{Non-equilibrium emergence}
Non-equilibrium emergence refers to the appearance of organized, collective phenomena in systems driven out of thermodynamic equilibrium \cite{raju2021diversity}. Unlike equilibrium phase transitions, where order arises through free energy minimization and spontaneous symmetry breaking in response to thermodynamic variables, non-equilibrium emergent behaviors are sustained by continuous energy or matter flows. These systems are inherently dissipative and rely on energy exchange with the environment to stabilize patterns, structures, or behaviors that cannot exist under equilibrium conditions. Consequently, their dynamics cannot be fully described by initial and final states alone, but depend on the ongoing interplay of driving and dissipation.
A paradigmatic example of non-equilibrium self-organization is the Rayleigh–Bénard instability in fluid mechanics \cite{berge1984rayleigh}. Here, a temperature gradient applied across a fluid layer causes buoyancy-driven convection, where warmer, less dense fluid drifts upwards and cooler, denser fluid sinks. This competition between buoyant forces and viscous dissipation gives rise to convection rolls once the Rayleigh number, a dimensionless parameter quantifying the relative strength of these forces, exceeds a certain critical threshold. Here, macroscopic ordered convective patterns emerge from local interactions between fluid elements as a result of the balance between energy input (heating from below) and dissipation (heat loss at the top). 
Non-equilibrium emergence is not limited to classical systems but extends to diverse physical and biological contexts. In soft matter physics, external mechanical stress can drive shear-induced instabilities in complex fluids such as emulsions and polymer melts. For example, shear band formation can occur in wormlike micellar solutions, which are complex fluids composed of self-assembled surfactant molecules \cite{fardin2016shear}. When subjected to shear flow, these micelles align and reorganize into distinct regions of low and high shear rates, leading to the emergence of shear bands. In this system, the non-linear coupling between the applied shear rate and the micellar alignment drives the fluid into a non-equilibrium steady state, where the molecular reorganization accommodates dynamically the external stress. As a result, spatially separated regions with different flow velocities and micelle ordering are formed \cite{helgeson2009rheology}. Flow-induced alignment in polymers, e.g., freeze-casting of colloidal suspensions, is another example of how mechanical energy input induces local ordering, which propagates across the system to form macroscopic patterns \cite{fan2018creating}. In this process, a liquid suspension containing particles is subjected to unidirectional freezing, where ice crystals grow and push the suspended particles into the regions between the advancing freezing fronts. This mechanical driving force induces local alignment of the particles, which propagates across the system, resulting in the formation of macroscopic, highly ordered porous structures. The anisotropic alignment created during freeze-casting gives rise to materials with tailored mechanical and functional properties that are widely used in applications like biomimetic scaffolds, thermal insulation materials, and filtration membranes \cite{lavoine2017nanocellulose}. 
In biological systems, non-equilibrium emergence underlies processes essential to life, where energy flows across multiple scales maintain dynamic order. For example, self-organization in cytoskeletal networks, driven by ATP hydrolysis, exemplifies how continuous energy consumption sustains structural patterns necessary for cellular function \cite{koehler2011collective}. 
Non-equilibrium emergence occurs also in the so-called open quantum systems, when continuous energy exchange between a quantum-mechanical system and an external quantum system (i.e., the environment) drives the system into novel steady-state configurations, that would not exist under equilibrium conditions, giving rise to dissipative phase transitions \cite{hwang2018dissipative}.
In a cavity quantum electrodynamics (QED) system, for example, quantum emitters (e.g., atoms, molecules, or artificial qubits) are strongly coupled to a confined electromagnetic field within an optical cavity. When the system is driven by an external field (e.g., a laser), the energy input compensates for the continuous photon loss (dissipation) through the cavity walls \cite{gleyzes2007quantum}. At a critical driving strength, the system undergoes a transition into a coherent state where the quantum emitters and the cavity field become synchronized, leading to the emission of coherent light, a phenomenon analogous to lasing. In this steady-state behavior, energy gain from driving fields competes with photon loss, and instead of destroying coherence, the dissipation stabilizes a long-range ordered quantum state. The resulting non-equilibrium steady state exhibits emergent order, such as macroscopic quantum coherence, that cannot be described using equilibrium thermodynamics. Therefore, dissipation, traditionally seen as a source of decoherence, can act as a stabilizing force for emergent order in quantum systems, under the right conditions. These principles are at the base of modern quantum technologies, where driving and dissipation are controlled to engineer desired quantum states far from equilibrium for applications in quantum computing, quantum sensing, and quantum communication \cite{harrington2022engineered}.

\subsection{Emergent multiscale structures}
Intrinsic emergence in multiscale structures arises in hierarchical systems, where multiple characteristic scales are intrinsically embedded in the structural organization of the system \cite{whitesides2002self}. These scales interact nonlinearly across different hierarchical levels, leading to emergent behaviors that originate in fluctuations that never become scale-invariant (unlike the phenomenology of phase transitions). A defining feature of such systems is interscale dissimilarity, for which the structural and functional characteristics at one scale differ qualitatively from those at another scale, but nonetheless remain interdependent through interscale coupling \cite{bagrov2020multiscale}. This interplay introduces structural and functional complexity that is in itself "emergent" as it cannot be captured through single-scale approximations.
Hierarchical materials, such as cellulose, bone, or spider silk, are typical examples of this intrinsic emergence. As seen in section \ref{self-assembled}, the characteristic dimensions of these systems span multiple orders of magnitude, from molecular to macroscopic scales, with each scale contributing unique, yet interconnected, information to the overall material behavior. The emergent properties arising from these multiscale interactions are predominantly mechanical (e.g., stiffness, toughness, elasticity), which reflect complex coupling and feedback across the different scales.
In cellulose, structural correlations begin at the molecular scale, where glucose monomers form polymer chains stabilized by covalent and hydrogen bonds. These chains provide local stiffness and stability but also serve as the building blocks for larger-scale organization. As the chains self-assemble at the nanoscale to form single microfibrils, they create alternating crystalline and amorphous domains characterized by different degrees of structural order. This is needed to balance rigidity and toughness in the final structure as crystalline domains contribute high tensile strength and stiffness, while amorphous regions introduce flexibility and energy dissipation \cite{rongpipi2019progress}.
At the mesoscale, microfibrils bundle into larger networks, where entanglement and interfacial interactions between fibrils allow for stress distribution and damage tolerance. These mesoscale networks are anisotropic, meaning that their mechanical response depends on the orientation of the fibrils relative to the applied force. For example, when fibrils are aligned along the direction of a load, stiffness and tensile strength are maximized, whereas an entangled, disordered arrangement enhances resilience to deformation.
Moving up to the microscale, these networks form fiber assemblies, which determine bulk material behavior and organize into larger networks at the macroscopic scale, defining the overall structural response to external forces. 
Therefore, structural constraints (e.g., fiber alignment) and deformations (e.g., bending of fibers) at one scale, propagate between neighboring scales affecting their mechanical dynamics. This non-linear feedback mechanism is responsible for emergent behaviors like strain stiffening, where the material becomes stiffer under increasing deformation, or energy dissipation, where hierarchical coupling increases damage tolerance \cite{van2019hierarchical}. In practical examples of materials made of cellulose like wood, the fibers are highly aligned along the the grain, which refers to the direction of the wood fibers orientation and growth rings. This alignment gives the wood exceptional stiffness and strength parallel to its grain. However, perpendicular to the grain, across the fibers' alignment, the material is much weaker and more prone to fracture propagation \cite{kretschmann2010mechanical}.

Similar concepts apply to bone, where mineralized collagen fibrils exhibit a multiscale architecture. At the molecular level, collagen molecules, composed of triple helices, provide intrinsic flexibility and tensile strength due to their polymeric structure and hydrogen bonding \cite{ramachandran1954structure}. These collagen molecules serve as a scaffold onto which nanoscale hydroxyapatite mineral platelets are deposited, providing rigidity and compressive strength. These platelets, which are oriented along the collagen fibrils, are nanometers thick and several tens of nanometers long, creating a composite material that exploits the mechanical properties of both components \cite{nair2013molecular}.
At the mesoscale, the mineralized collagen fibrils bundle into networks, where the interplay between the compliant organic matrix (the collagen) and the stiff inorganic phase (the mineral platelets) introduces non-linear mechanical responses to preserve its structural integrity. The fibrils interact through sliding mechanisms, where energy is dissipated during deformation, preventing catastrophic failure under loading.
Moving up to the microscale, these fibril networks assemble into lamellar structures, within compact bone, or trabecular struts, within spongy bone. In trabecular bone, the open-cell architecture further contributes to bone’s mechanical properties by efficiently redistributing stresses. Due to this hierarchical design, as a crack propagates through the bone, its path is redirected along interfaces between mineral platelets and collagen fibrils, dissipating energy and delaying fracture. Emergent properties resulting from this multiscale interaction are damage tolerance, toughness, and resistance against both compressive and tensile stresses \cite{Fratzl2007Bone}.

In spider silk, nanoscale molecular alignment within fibrils provides tensile strength, while mesoscale fibril packing enables controlled energy dissipation in response to deformation. At the macroscopic scale, the woven arrangement of spider silk fibers allows for extensibility by accommodating deformation through fiber reorientation and sliding, while also enhancing resilience by distributing stresses across the network. The emergent property of this hierarchical design would be elastic recovery, which prevents localized failure as it enables the silk to deform under tensile loading and redistribute the mechanical stress throughout its structure \cite{gosline1999mechanical}.

Beyond emergent properties of mechanical kind, hierarchical structures exhibit also emergent optical, and transport (acoustic, thermal, electrical) properties. For example, butterfly wings exhibit truly multifunctional emergent properties by combining nanoscale optical effects, mesoscale layering, and macroscopic lightweight design. The surface of butterfly wings at the molecular level is composed of chitin, a biopolymer similar to cellulose that provides structural stability through hydrogen bonding. At the nanoscale, chitin organizes into photonic crystal-like structures with periodic arrangements that interact via interference, diffraction, and scattering with visible light. Here, vibrant, angle-dependent iridescent colors emerge without pigments \cite{poladian2009iridescence}. At the mesoscale, the photonic crystal nanostructures integrate into patterns of layered squamae, that coat the wings and reflect infrared light while selectively absorbing other wavelengths, which allows for efficient thermal regulation of the butterfly's body. At the microscale, the squamae overlap like shingles, enhancing the wing’s resilience to mechanical damage while minimizing weight (see Fig. \ref{miller}b). At the macroscale, butterfly wings exhibit emergent properties such as lightweight stiffness and flexibility, that allow the wings to withstand aerodynamic forces during flight.

Unlike equilibrium phase transitions, where emergent properties arise through critical phenomena near equilibrium, the emergent behavior in hierarchical systems is intrinsic and persists independently of external thermodynamic conditions. Unlike non-equilibrium emergence, which arises from continuous energy or matter flows driving the system far from equilibrium, hierarchical systems possess pre-existing structural organization across multiple scales originating in self-assembly \cite{whitesides2002self}. These intrinsic scales interact in a non-linear fashion, giving rise to emergent properties that are present even in static or near-equilibrium conditions. However, when hierarchical systems are driven far from equilibrium (e.g., through mechanical deformation) their multiscale structure couples dynamically with dissipative processes, which enables adaptive responses \cite{kretschmann2010mechanical}. In this sense it could be said that hierarchical systems bring the two worlds together, since static multiscale organization and non-equilibrium dynamics coexist, with the multiple pre-existing scales contributing simultaneously to the system's response. This is ultimately the reason why conventional single-scale methods, traditionally implemented for periodic structures, fail to capture this complexity: they do not account for the non-trivial interactions between scales.

\section{Characterizing emergence and structural order}
Experimental methods for the characterization of condensed matter systems can be broadly categorized into those focusing on structural organization based on scattering and imaging techniques, and those targeting dynamical processes or response properties, including spectroscopy (at microscopic length scales) and macroscopic measurements like thermal transport, mechanical testing, resistance and magnetization \cite{chaikin1995principles}. Here, structure and dynamics can be probed at different length and time scales but to clarify microscopic origin of emergent phenomena, spectroscopic methods are particularly powerful, as they probe the elementary excitations (e.g., phonons, magnons, and electronic transitions) that can be directly connected to the fundamental interplay between structure and properties \cite{ashcroft1978solid}. In periodic systems, such as crystalline solids, this relationship is relatively straightforward: the well-defined arrangement of atoms and translational symmetry allow for a bottom-up approach, where emergent properties can be derived from microscopic organization. Diffraction techniques, reciprocal space analysis, and inelastic scattering are the bedrock of this characterization framework, as they have allowed to map structure-property relationships at single scales for decades. Systems that are not easy to standardize, including soft matter, amorphous materials, and hierarchical structures, pose challenges for characterization which are mainly due to their lack of long range periodicity, as mentioned in section \ref{order}. For such systems, the traditional methods developed for periodic systems remain foundational tools for exploring structure-property relationships, but their application needs to be adapted to the specific needs of the investigated system \cite{cross2015materials}. Scattering techniques, originally designed to exploit long-range periodicity, can also extract correlations in disordered or partially ordered materials due to the broad applicability of the Fourier transform. Real-space imaging methods are now sufficiently advanced to provide direct visualization of local and mesoscale organization in high resolution \cite{\cite{sakdinawat2010nanoscale}}. Vibrational and dynamical measurements, once primarily focused on phonons and electronic excitations in periodic systems, are now explored to probe dynamics in disordered and soft matter systems \cite{wang2024recent}.

In this section we discuss how single-scale characterization methods are applied, extended, or reinterpreted to investigate structure-property relationships in both periodic and non-standard systems. 

\subsection{Elastic Scattering Methods} 
Wide-angle and small-angle scattering techniques are used to characterize the average static structural organization of materials on the atomic to mesoscopic length scales, via elastic scattering of probe particles with scattering centers constituted by single atoms in one case and particle surfaces or interfaces in the other \cite{hamley2021small}. In practice, during a scattering experiment we measure the intensity of the scattered waves resulting from this interaction as a function of the scattering vector $\boldsymbol{q}$, which is defined as the difference between the incident and scattered wave vectors of the probing radiation: $\boldsymbol{q} = \boldsymbol{k}_{out} - \boldsymbol{k}_{in}$. Since in elastic scattering the moduli of the wave vectors remain unchanged, the value of $\lvert \boldsymbol{q} \rvert$ can be expressed as a function of the scattering angle 2$\theta$ (i.e., the angle formed between $\boldsymbol{k}_{in}$ and $\boldsymbol{k}_{out}$), from a simple trigonometric argument: $q = \frac{4\pi}{\lambda} \sin{\theta}$. By combining this expression with the Bragg's law in equation \ref{bragg}, it is easy to see that there is a relation of inverse proportionality between $\boldsymbol{q}$ and the real-space characteristic size of the measured feature $d$, given by $d\sim 2\pi/q$. Therefore, large $q$ values correspond to small real-space lengthscales ($\sim$ Å) comparable to atomic distances, which produce scattered intensities at wide angles. Small $q$ values correspond to larger real-space scales ($\sim$ 10$^2$ nm), comparable to the size of macromolecules, nanoparticles, aggregates or mesoscopic domains (e.g., pores), which produce scattered intensities at small angles.
Interestingly, the nomenclature wide-angle scattering is commonly used in soft-matter physics frameworks, but the very same technique is known as diffraction in solid-state physics \cite{gregory1957elements}. Here, the position and relative intensities of Bragg peaks in the diffraction pattern determine the structural organization of crystalline materials. For single crystals, diffraction experiments enable direct access to reciprocal space coordinates by physically rotating the sample to satisfy constructive interference conditions, thereby mapping of the full 3D reciprocal lattice \cite{facio2023engineering,thomarat2024tuning}. This information facilitates the determination of lattice parameters, space groups, and atomic arrangements through symmetry analysis and automated refinement techniques \cite{sheldrick2015shelxt,rodriguez2001introduction}. In powder diffraction, where isotropic diffraction rings are produced due to the averaging over all possible crystallite orientations \cite{matsubara2020cation,matsubara2020magnetism}, Bragg peaks are indexed, and unit cell parameters are deduced using semi-automatic algorithms. The atomic structure is then retrieved using methods like the Rietveld refinement, which models the scattering profile and minimizes the weighted difference between observed and calculated intensities \cite{rietveld1969profile}. In soft matter there is a need to differentiate wide from small angle scattering (SAS), as the two techniques are often used together to probe simultaneously a broader range of lenghtscales in a single measurement. This is the case for partially ordered/hierarchical materials (e.g., block copolymers, colloidal suspensions, biomaterials) where the combination of SAS and wide-angle scattering measurements have revealed how nanoscale crystalline domains coexist with amorphous regions, producing broad spectral features that reflect overlapping contributions from multiple scales \cite{Nishiyama2002Cellulose,cherny2014small,schiele2024influence,aahl2025multimodal}.
Of course, the Bragg's law itself was created in a framework where the scattering target could be treated as a periodic grating made of parallel planes. In non-periodic systems, it does not always make sense to discuss in terms of lattice planes and the interference patterns arise from more general principles related to the Fourier transform of spatial correlations between scattering centers \cite{hansen2013theory}. Specifically, scattered waves interfere based on short-range correlations or local density fluctuations. The interference pattern is continuous rather than discrete, producing diffuse scattering with broad features. Nonetheless, the relationship between $q$ and $d$ remains valid because the scattering intensity still corresponds to the Fourier transform of the spatial density distribution $\rho (\boldsymbol{r})$ according to the relationship:

\begin{eqnarray}
I(q) \propto \lvert \int \rho(\boldsymbol{r}) e^{-i \boldsymbol{q} \cdot \boldsymbol{r}}) d\boldsymbol{r} \rvert ^{2}.
\label{iq}
\end{eqnarray}

Here, $\rho(\boldsymbol{r})$ represents the real-space material's density. Features in $I(q)$ correspond to characteristic real-space distances $r \sim 2\pi/q$, even without periodicity. Pair Distribution Function (PDF) analysis \cite{wang2020pair} transforms the scattering intensity into real-space correlation functions $G(r)$, from which it is possible to determine average nearest-neighbor distances, or detect local structural motifs \cite{brant2021pressure}. In soft matter systems the elementary components of the structure (i.e., the particles) will have different shapes depending on the type of system investigated. Therefore, in realistic SAS experiments the total density distribution $\rho(\boldsymbol{r})$ includes two different contributions: $\rho_{particle}(\boldsymbol{r})$, which describes the density distribution within a single particle, and $\rho_{interparticle}(\boldsymbol{r})$, which accounts for the spatial arrangement of multiple particles \cite{hamley2021small}. Therefore, the Fourier transform in equation \ref{iq} yields the following relationship:

\begin{eqnarray}
I(q) = N V^{2} \Delta \rho^{2} P(q)S(q),
\label{structureform}
\end{eqnarray}

where $N$ and $V$ are the number and volume of particles in the system respectively, $\Delta \rho$ is the contrast in scattering length density between the particle and the surrounding medium. The latter parameter is a measure of how strongly a material or a region within a material interacts with the incident radiation in a scattering experiment. $P(q)$, called the form factor, is the Fourier transform of $\rho_{particle}(\boldsymbol{r})$ and describes the scattering due to the intrinsic structure of the particle. $S(q)$, the structure factor already mentioned in section \ref{Reciprocal space in soft matter}, is actually the Fourier transform of $\rho_{interparticle}(\boldsymbol{r})$ and describes the spatial correlations between particles. 
Therefore in SAS, the scattering intensity at high $q$ is dominated by the $P(q)$ term, while at intermediate to low $q$ the $S(q)$ term is dominant. If the system is highly diluted, $S(q)$ = 1 and the scattering intensity will only depend on $P(q)$. The profile of $P(q)$ depends on the particle's size and shape, and its functional form is well-documented for common geometries such as spheres, cylinders, and disks. By way of example, in block copolymers, $P(q)$ reflects the internal structure of individual domains such as lamellae or micelles, while $S(q)$ reflects the periodicity and long-range order in the self-assembled patterns \cite{hamley2004small,}. 

Once the contributions from these two terms are disentangled, SAS in combination with wide-angle scattering can provide a comprehensive view of structural organization across length scales up to hundreds nanometers. Given their ability to probe both local and long-range correlations, these techniques are essentially the most reliable and widely applicable structural characterization methods for soft matter systems. However, the interpretation of SAS data is challenging due to the complexity and variability of the structures involved. Unlike crystalline systems, whose symmetry constraints allow straightforward structural refinement using semi-automatic algorithms, SAS cannot directly yield "the structure" of a system simply through data fitting. This is because the scattering profiles in SAS often exhibit exponential decay trends, with little to no features, where key structural characteristics need to be identified \textit{a-priori} to guide the fitting process \cite{petoukhov2021ambiguity}. Once the specific scattering regimes have been identified, the desired structural information can be extracted by applying the appropriate quantitative model. 

Many models have been developed to address the diversity of systems encountered in SAS, each of them adapted to extract specific types of information from the scattering profile. To mention a few important examples: Guinier analysis is a fundamental approach for extracting the radius of gyration of particles or clusters, which allows to determine their overall size and shape in the low-$q$ regime, where the scattering intensity is dominated by the largest structural features \cite{guinier1956small}. Porod analysis instead is widely used in the high-$q$ regime, to investigate surface and interface properties, where the scattering intensity follows a characteristic power-law decay that reflects the smoothness or roughness of the interfaces between structural elements \cite{kratky1949diffuse}. Fractal models are used for systems with self-similar structures, such as aggregated colloids or porous networks; these models allow to extract the mass fractal dimension (which describes how mass is distributed) or the surface fractal dimension (which quantifies surface roughness) from the slope of the power-law trend of the scattering profile \cite{anitas2020small}. Correlation peak analysis is used for studying systems with well-defined characteristic length scales, such as the interparticle spacings in colloidal suspensions or the periodic domain spacing in block copolymers, as these features manifest as distinct peaks in the scattering profile \cite{hamley2004small}. Polydispersity models account for heterogeneous samples, with a distribution of particle sizes, where the observed scattering intensity reflects the weighted contributions of such different sizes \cite{kotlarchyk1983analysis}.
The choice of the model relies on prior understanding of the system under investigation and depends on its specific structural characteristics, as well as on the range of the scattering vector $q$ probed in the experiment. As structural contributions may overlap in the same material, a combined approach using multiple models is often needed when investigating complex systems. However, analytical models typically assume idealized shapes (perfect spheres, rods, lamellae), uniform scattering length densities, and simple cases for interparticle interactions. In reality, many systems deviate from these assumptions and simulations, such as Monte Carlo or molecular dynamics, are also often needed to complement the experimental data as in, e.g., protein structures \cite{lycksell2022biophysical,lycksell2021probing}.

\subsection{Real Space Imaging}
Real-space imaging incorporates a broad range of techniques that are able to map a sample’s structure directly in real (physical) space providing spatially resolved images of morphology, topology, or local composition. Unlike reciprocal-space techniques, that require mathematical transformation to infer structure, real-space imaging generates spatially resolved images that can be directly interpreted. The imaging process works by recording the interaction of a probe (i.e., photons, electrons, neutrons, or mechanical tips) with the material, and mapping these interactions into image contrast based on underlying physical properties. Here, resolution and contrast are, in fact, the fundamental parameters that directly impact the ability to extract meaningful information from images \cite{trianni2025physics}. Resolution refers to the ability of the imaging system to differentiate between two closely spaced features, which translates in the smallest spatial detail that can be distinguished in the resulting image, and it depends on the wavelength of the probing radiation or probe size. Contrast mechanisms refer to how differences in a sample's properties create variations in the image intensity, which allows to visually highlight different structural features. These properties can be atomic number (for electron microscopy), local height/topography (for scanning probe methods), refractive index (for optical microscopy), or chemical affinity (for certain spectroscopic imaging techniques). Therefore, depending on the type of sample and the specific research needs, various imaging methods are employed, each relying on distinct physical principles of probe-matter interaction to generate contrast, which can take place through different mechanisms. Specifically, phase contrast occurs due to differences in the phase shifts of the probe wavefront, as it goes through regions of varying thickness or refractive index within the sample, which makes it suitable for visualizing subtle internal structures. Absorption contrast is based on differences in the attenuation of the probe as it passes through the material; this mechanism is well-suited for highlighting variations in density or composition. Finally, scattering contrast arises from interactions between the probe and structural heterogeneities in the sample, creating localized intensity variations that reveal features like grain boundaries, defects, or interfaces.

\subsubsection{Optical microscopy (OM)} is the earliest form of microscopy. It uses visible light to produce magnified images of an object through a set of lenses, and is traditionally constrained by the diffraction limit, which restricts resolution to approximately half the wavelength $\lambda$ of light ($\sim$200 nm) \cite{mertz2019introduction}. This limit is due to the wave nature of light, where diffraction blurs out closely spaced features and is defined by the Rayleigh criterion:

\begin{eqnarray}
d = 0.61 \frac{\lambda}{NA},
\label{ray}
\end{eqnarray}

where NA is the numerical aperture of the lens of the microscope and 0.61 is a dimensionless scaling factor derived from the diffraction physics of light and the geometry of circular apertures.
However, advancements such as confocal microscopy, which uses point illumination and pinhole apertures to reject out-of-focus light, and super-resolution techniques like STED (stimulated emission depletion) and PALM/STORM (photoactivated localization microscopy), have pushed the resolution of OM well below the diffraction limit. These methods enable imaging of subcellular structures and nanoscale features, which is particularly relevant for biological and soft materials.

\subsubsection{Electron microscopy (EM)} is a method in which a beam of high-energy electrons interact with the sample's atomic structure through scattering and absorption. Here, due to the wave-like behavior of electrons, these interactions are governed by relativistic electron wavelengths described by the de Broglie equation $\lambda = \frac{h}{p}$, where $h$ is the Planck's constant and $p$ is the electron momentum. As electrons are accelerated to near-light speeds, their momentum increases and their wavelength is reduced to values much smaller than visible light, which allows EM to achieve resolutions down to the atomic scale.
Depending on the specific imaging needs, specialized forms of EM can be adopted. Scanning Electron Microscopy (SEM) scans a focused electron beam across a sample surface, detecting secondary or backscattered electrons to create high-resolution topographical or compositional maps \cite{goldstein2017scanning}. Transmission Electron Microscopy (TEM), instead, transmits electrons through an ultra-thin sample, producing detailed images of internal structures at sub-nanometer resolution and is suitable for atomic-scale imaging of crystalline materials, interfaces, and defects \cite{carter2016transmission}. 
Both these traditional methods face significant limitations when imaging biological samples or soft materials, as the high-energy electron beams can cause radiation damage, structural degradation, or desiccation of hydrated samples in the vacuum environment \cite{glaeser1978radiation}. In fact, to mitigate these effects, measurements are typically performed under short acquisition times \cite{willhammar2021local}. Recent advancements, particularly in TEM, try to address these challenges by introducing cryo-TEM, a novel technique that operates by freezing the materials in vitreous ice, which should preserve their near-native states and prevent water evaporation \cite{milne2013cryo}. By using low-dose electron beams, cryo-TEM also minimizes radiation damage, which allowed to study proteins, viruses, and lipid membranes with atomic precision.

\subsubsection{Scanning Probe Microscopy (SPM)} \cite{bian2021scanning} measures physical forces (e.g., van der Waals, tunneling current) between a sharp probe and the sample's surface, producing high-resolution topographic or electronic maps through scanning. Specific types of SPM techniques are Scanning tunneling microscopy (STM) \cite{chen2021introduction} and atomic force microscopy (AFM) \cite{voigtlander2015scanning}. STM operates on the principle of quantum tunneling, a phenomenon where electrons pass through the vacuum gap between a sharp metallic probe and the surface of a conductive sample when the separation is small (a few Å). By controlling the distance between the probe and the sample, STM measures the tunneling current $I$, which is highly sensitive to the separation at the atomic scale and follows an exponential decay trend of the type: 

\begin{eqnarray}
I \propto e^{-2kd},
\label{stm}
\end{eqnarray}

where $d$ is the distance between the sample and the tip and $k$ is a decay constant related to the work function of the material. The tunneling probability, and thus the current, decreases exponentially with increasing distance $d$. This sharp dependence allows the STM to achieve sub-nanometer spatial resolution, while the dependence on $k$ governs the sensitivity of the tunneling current to the materials that constitute the tip and sample. This tunneling current is directly related to the local electron density of the sample surface, allowing STM to generate high-resolution, three-dimensional maps of the surface's electronic structure and states. STM is therefore capable of visualizing individual atoms and also provides insights into the electronic properties and behaviors of the measured material at the nanoscale, yielding direct insights into phenomena like superconductivity or charge density waves \cite{yuan2024superconductivity,fang2020robust}. AFM utilizes a cantilever with a sharp probe tip to detect and measure interatomic forces between the tip and the sample surface. These forces include van der Waals forces, electrostatic forces, and other interactions, depending on the operating mode of the instrument. As the tip scans the surface, the cantilever is deflected in response to these forces, and the deflection is typically measured using a laser beam reflected off the back of the cantilever onto a photodetector. This allows AFM to construct highly detailed, three-dimensional images of the sample’s topography. Unlike STM, AFM is not limited to conductive samples and its ability to operate in different modes, such as contact, non-contact, and tapping, makes it more versatile for utilization on diverse sample types \cite{last2010applications}. Beyond topography, SPM techniques can measure local mechanical, electrical, and thermal properties, offering the possibility to study nanoscale surface emergent phenomena beyond the mere structure, as mentioned above. However, these methods have limitations, as they are primarily surface-sensitive and cannot really probe subsurface structures. They also require smooth, well-prepared samples, and some modes are restricted to specific types of materials (e.g., conductive samples for STM). Fragile or highly soft samples, like certain biological tissues, may suffer from deformation or damage during scanning, while large or highly irregular surfaces and multi component systems can pose challenges for accurate imaging \cite{joshua2023soft}. 

\subsubsection{X-ray and Neutron imaging} \cite{martz2016x,strobl2024neutron} operate on the principle of beam attenuation, which is mathematically described by Beer-Lambert’s law:

\begin{eqnarray}
I = I_0 e^{-\mu x},
\label{beer}
\end{eqnarray}

where $I_0$ is the intensity of the incident beam, $I$ is the transmitted intensity after passing through a material of thickness $x$, and $\mu$ is the linear attenuation coefficient, which characterizes the material’s interaction with the probing radiation.  While the exponential form of Beer-Lambert’s law remains valid for both X-ray and neutron imaging, the physical mechanisms underlying attenuation differ. 

For X-rays, attenuation occurs primarily due to photoelectric absorption and Compton scattering, and $\mu$ is largely determined by the material’s atomic number $Z$ and density. High-$Z$ materials strongly absorb X-rays, while low-$Z$ materials, such as organic compounds, are more transparent. The resulting attenuation provides contrast in the produced image, displaying bulk structural features based on differences in absorption characteristics of the various components within the sample \cite{martin2007importance}.
For neutrons, attenuation is governed by nuclear interactions, including elastic and inelastic scattering, as well as absorption. Unlike X-rays, neutron attenuation is highly dependent on elemental and isotopic composition, rather than atomic number. Some elements, like hydrogen and boron, exhibit strong neutron absorption, while others, like lead, are relatively transparent. Consequently, neutrons have a significantly higher penetration capability with respect to X-rays, and the resulting contrast in the produced image is sensitive to isotopic variations and elements which interact weakly with X-rays \cite{kardjilov2011neutron}.

X-ray and neutron imaging techniques are therefore complementary and both capable of probing a wide range of length scales, from millimeter to micrometer-sized bulk features, but advances in high-resolution detectors and synchrotron-based X-ray imaging have pushed spatial resolution for X-ray imaging methods into the nanoscale \cite{sakdinawat2010nanoscale}. 

These methods also overcome the limitations of EM and OM by addressing major gaps in imaging depth, contrast, and the ability to non-destructively study the internal structure of bulk materials. In-fact, as EM techniques are fundamentally limited to surface-sensitive studies or ultra-thin specimens, they cannot investigate the internal structures of large, opaque samples without significant sample manipulation, which inevitably alters the material. Similarly OM, though convenient and widely accessible, is constrained by its low resolution and limited penetration depth, which does not allow to see through dense and opaque materials that absorb visible light.
For this reason, since X-rays and neutrons allow bulk imaging due to their deep penetration capabilities into matter, they are considered non-destructive. 

These bulk characterization capabilities remarkably enable tomography, that reconstructs 3D representation of a sample's internal structure from a series of 2D projection images taken at different angles. The fundamental principle underlying tomography is the Radon Transform, which relates these projections to line integrals of attenuation within the sample \cite{radon1986determination}. Using computational algorithms, such as Filtered Back Projection or iterative methods, the inverse Radon Transform reconstructs the internal distribution of attenuation coefficients in 3D \cite{gordon1974three}. The result is a volumetric dataset that contains the sample's internal features such as cracks, pores, or phase boundaries, which are directly visible through virtual cuts of the reconstructed image. It should be mentioned that, while tomography is widely used in X-ray and neutron imaging, it is also possible to some extent with electrons, where a transmission electron microscope captures tilted 2D projections of ultra-thin samples to achieve nanoscale or even atomic-resolution 3D reconstructions \cite{ercius2015electron}. X-ray and neutron imaging (especially neutrons) typically require access to large-scale facilities such as synchrotrons or neutron sources, which is impractical with respect to utilizing in-house and table-top methods like OM and EM. Nonetheless, the investment in these resources is justified by their ability to answer critical scientific and engineering questions that in-house methods cannot address. Large-scale experimental facilities enable the real-time study of materials under extreme conditions (e.g. high/low temperature, pressure, or stress) or the in-situ and dynamic study of processes such as fluid flow, crack propagation, and phase transitions \cite{shafabakhsh20244d,liu2011situ,ramli2022liquid}.

However, these techniques also have their limitations. For X-rays is challenging to image low-contrast features in materials composed of elements with similar atomic numbers, which requires specialized methods like phase contrast or high-energy imaging to enhance visibility \cite{wang2019high}. Radiation damage is another critical consideration for high-energy X-rays, that can deteriorate sensitive materials through ionization and heating. Ionization occurs when X-ray photons transfer enough energy to the atoms in the probed material to eject tightly bound electrons, creating charged ions and free electrons. This process can break chemical bonds, leading to structural degradation or changes in the material's chemical composition \cite{holton2009beginner}. Heating occurs when the X-ray energy deposited into the material is converted into heat (i.e., thermal motion). This localized increase in temperature can cause physical changes such as melting, thermal expansion, or even destruction of samples that are temperature-sensitive, such as biological specimens or certain polymers. Both effects are cumulative and must be minimized during imaging, often by reducing the X-ray dose or cooling the sample to mitigate damage. Neutron imaging is effective for the characterization of hydrogen-rich and isotopically distinct samples, and does not pose critical concerns in terms of radiation damage. However, neutron imaging typically has lower spatial resolution than X-rays due to the need for indirect detection, where neutrons must interact with a converter material (i.e., a scintillator) to produce detectable signals, leading to spatial spreading \cite{hussey2021high}. Additionally, neutron beams have a relatively low flux compared to X-rays and thicker scintillators and lower pixel densities, needed to enhance neutron capture efficiency, further limit the imaging resolution.

In general, imaging methods offer access to a relatively broad range of length scales compared to scattering methods, as they can be adapted to visualize structures from the nanoscale to the macroscale. However, unlike scattering methods, which provide averaged information over multiple length scales simultaneously, imaging techniques typically focus on a specific length scale at a time, requiring adjustments to access a different scale. Therefore, they also face fundamental shortcomings when addressing the challenge of multiscale characterization, which ultimately requires simultaneous observation of features across orders of magnitude in length scales. Most imaging techniques deal with a trade-off between spatial resolution and field of view, which defines the spatial region visible within the imaging system. High-resolution methods are confined to ultra-thin samples and small regions of interest, which interdicts their applicability to bulk materials or systems with macroscale dimensions. On the other hand, methods that are suitable for larger-scale imaging lack the resolution required to probe finer structural details at the nano and sub-nano scale, leaving critical subtle features unresolved. Moreover, even though real-time measurements of dynamic processes are possible, they are often constrained to either bulk or small-scale phenomena. Observing the interplay between processes occurring at vastly different scales, such as atomic-level reactions driving macroscale structural changes, remains a significant challenge. For this reason, nowadays, modern workflows often integrate multiple imaging modalities in the same sample region to capture different aspects of structures at relevant length scales \cite{correlative_microscopy,nygaard2024formax}, Advanced reconstruction algorithms have evolved significantly, transitioning from traditional methods like filtered back projection to contemporary approaches incorporating artificial intelligence, thereby enhancing image quality and reducing artifacts \cite{willemink2019evolution}. Such multimodal approaches yield quantitative morphological data that can be correlated directly with scattering or spectroscopic results.


\subsection{Spectroscopic Methods}

Unlike elastic scattering and real-space imaging, which primarily reveal static structures, spectroscopic techniques probe dynamical excitations in condensed matter, providing insights into how materials respond to perturbations and how their structural organization influences emergent properties like charge and heat transport, spin dynamics, and optical behaviors \cite{chaikin1995principles}. A fundamental class of such excitations is phonons, which represent the quantized vibrational modes of a material’s atomic structure. Phonons are present in all condensed matter systems as every material, whether crystalline or not, exhibits some form of structural vibration, though the nature of these vibrations depends on structural order \cite{ashcroft1978solid}. In crystalline solids, the periodic arrangement of atoms enables phonons to be well-defined quasiparticles, propagating through the lattice with a characteristic energy–momentum dispersion relation, $E(\boldsymbol{Q})$ (or also $\omega(\boldsymbol{Q})$), that can be directly measured with conventional spectroscopic techniques \cite{nocerino2023q}. The same periodicity also allows for the emergence of other well-defined collective excitations in systems where electronic states exhibit long-range coherence (e.g., semiconductors, metals, and strongly correlated quantum materials), such as magnons (quanta of spin waves in magnetic materials) \cite{portnichenko2016magnon} and electronic bands, which describe the allowed energy states of electrons \cite{horio2018three}.

Hierarchical and amorphous systems lack translational symmetry and are inherently multiscale, which makes the conventional notion of phonons ill-defined (typically referred to as “phonon-like” or “quasi-phonon” excitations) and introduces specific characterization challenges. Due to the absence of atomic periodicity, excitations cannot be straightforwardly mapped in reciprocal space, making traditional dispersion analysis inapplicable \cite{mu2020unfolding}. In these systems, multiple length/time scales, and their dynamic interplay, contribute simultaneously to the measured response which, together with structural complexity, lead to broadened spectral features, multiple scattering effects, and complex energy dissipation pathways. Consequently, vibrational modes in hierarchical materials may exist only transiently or over limited length scales (strong damping), become spatially confined within specific regions (localized modes), or hybridize over multiple hierarchical scales (delocalized modes). While phonon modes in crystalline solids are typically delocalized but well-described by dispersion relations, in hierarchical and disordered systems, these short-lived, hybridized vibrational interactions require alternative frameworks for analysis and interpretation \cite{mehra2018thermal}.

Given the diversity of condensed matter systems, a wide range of spectroscopic techniques has been developed, each optimized to probe specific types of excitations across different length and time scales. In crystalline solids, where quasiparticles such as phonons and magnons are well-defined, reciprocal-space probes like inelastic neutron and X-ray scattering directly measure dispersion relations and excitation lifetimes. However, in hierarchical and disordered systems, where excitations lack long-range coherence, real-space and broadband spectroscopies such as Raman, terahertz, and Brillouin spectroscopy become relevant for extracting vibrational dynamics. In cases where electronic structure and relaxation processes dominate, time-resolved and ultrafast spectroscopies such as pump-probe and free-electron laser techniques provide direct access to transient excitations. Furthermore, local probes like NMR and muon spin rotation offer detailed insights into site-specific interactions and fluctuations in structurally complex materials. This section discusses these techniques in detail, outlining their working principles, conventional applications, and adaptations required for characterizing hierarchical and disordered systems.



\subsubsection{Infrared (IR) spectroscopy} measures the absorption of infrared radiation by molecular vibrations \cite{stuart2004infrared}. When infrared light interacts with a material, specific vibrational modes of molecules absorb energy, leading to characteristic absorption peaks in the IR spectrum. The absorption follows Beer-Lambert’s law on the form: $A = \epsilon cl$, where the absorbance $A$ is related to the concentration of the target material $c$, the path length of the radiation $l$ and the molar absorptivity $\epsilon$. For a vibrational mode to be IR-active, it must induce a change in the dipole moment of the molecule. The fundamental vibrational frequencies of molecular bonds are given by:

\begin{eqnarray}
\nu = \frac{1}{2\pi} \sqrt{\frac{k}{\mu}},
\label{IR}
\end{eqnarray}

where $k$ is the bond force constant and $\mu$ is the reduced mass of the vibrating atoms \cite{stuart2004infrared}. This means that IR spectroscopy is particularly sensitive to polar functional groups, such as O-H, C=O, and N-H stretching vibrations. Traditionally, IR spectroscopy has been extensively applied to solid-state materials, where it is used to identify molecular and collective lattice vibrations as well as phase transitions in crystalline solids. Inorganic materials and minerals exhibit phonon modes in the far-IR region, while organic and polymeric materials display distinct vibrational fingerprints in the mid-IR \cite{ozaki2021infrared}. Additionally, techniques such as attenuated total reflectance (ATR-IR) have extended its applicability to thin films and interfaces by enhancing surface sensitivity \cite{laroche2013ftir}. 

In well-ordered crystalline materials, IR spectra consist of sharp, well-defined peaks corresponding to distinct vibrational states, facilitating straightforward spectral interpretation. However, when applied to hierarchical and disordered systems, IR spectroscopy encounters several challenges. The loss of long-range order due to variations in bond lengths and angles, results in broadened vibrational features that make peak assignments less distinct. Furthermore, hierarchical materials often exhibit multiple chemically distinct environments, leading to overlapping absorption bands that obscure spectral interpretation. The heterogeneity of these systems also introduces multiple scattering effects, which can distort baseline stability and affect signal intensity. Additionally, anisotropic molecular arrangements, common in soft matter and biomaterials, require polarization-sensitive techniques to resolve orientation-dependent vibrational modes.

To address these challenges, various adaptations of IR spectroscopy have been developed to enhance spatial resolution and improve spectral interpretation. Hyperspectral IR imaging, for example, combines spectral and spatial resolution, to map the chemical environment across the investigated material \cite{roggo2005infrared}. Fourier-transform infrared spectroscopy (FTIR) improves signal-to-noise ratios and enables time-resolved studies of dynamic processes \cite{griffiths1977recent}. Near-field IR spectroscopy, which integrates IR with atomic force microscopy (AFM-IR), achieves nanometer-scale spatial resolution, making it possible to probe local vibrational properties in heterogeneous materials \cite{dazzi2017afm}. Infrared spectroscopic ellipsometry is particularly useful for analyzing layered structures and thin films, as it measures the change in polarization state of IR light upon reflection from a sample surface, which allows to extract quantitative information on film thickness, optical anisotropy, and interfacial bonding \cite{roseler2006infrared}.

\subsubsection{Raman spectroscopy} is a vibrational spectroscopic technique that probes molecular and lattice vibrations by measuring inelastically scattered light \cite{smith2019modern}. When monochromatic light, typically from a laser, interacts with a material, most photons undergo elastic (Rayleigh) scattering, where their energy remains unchanged. However, a small fraction of photons exchange energy with the vibrational modes of the material, resulting in a shift in wavelength known as the Raman shift. The energy difference between the incident and scattered photons corresponds to the vibrational energy levels of the material $\Delta E = h(\nu_i - \nu_s)$, where $\nu_i$ and $\nu_s$ are the frequencies of the incident and scattered photons respectively. These Raman shifts provide a fingerprint of molecular bonds, crystal structures, and phonon modes. The Raman intensity is proportional to changes in the material’s polarizability, making it complementary to infrared (IR) spectroscopy, which probes dipole-active vibrations.

In solid-state physics, Raman spectroscopy is traditionally used to investigate phonon dispersions, phase transitions, and symmetry-breaking effects in ordered materials. Here, well-defined selection rules, determined by group theory, restrict which vibrational modes are Raman-active, resulting in sharp and well-resolved peaks in the measured spectrum \cite{wu2024symmetry}. The technique is particularly valuable for studying structural distortions, strain effects, and electron-phonon coupling in semiconductors, superconductors, and low-dimensional materials (e.g., graphene and transition-metal dichalcogenides) \cite{cong2020application}. Resonant Raman scattering further enhances sensitivity to specific electronic transitions, enabling studies of excitonic and charge-transfer phenomena \cite{cirera2022charge}.

Hierarchical and disordered materials introduce several complexities that challenge conventional Raman spectroscopy. The loss of translational symmetry in these systems leads to a breakdown of strict selection rules, causing a broadening of spectral features and making it difficult to distinguish individual vibrational modes. Structural heterogeneity results in additional vibrational modes not predicted by symmetry considerations, complicating spectral interpretation. In nanostructured and amorphous materials, phonon localization and surface effects further modify phonon dispersions, leading to Raman shifts that locally depend on particle size, surface chemistry, and interface interactions. Moreover, in hierarchical systems where multiple length scales contribute to vibrational dynamics, Raman spectra can become highly complex because of mode hybridization between localized and delocalized phonon modes. Finally, multiscale interactions introduce multiple scattering effects, where vibrational energy is redistributed across different structural domains, obscuring clear assignments of Raman-active modes.

To address the challenges of Raman spectroscopy in complex systems, several advanced techniques have been developed to resolve vibrational signatures with enhanced spatial, spectral, and temporal resolution. Spatially resolved and confocal Raman microscopy, for example, utilizes a tightly focused laser spot ($\sim$300–500 nm) to enable real-space mapping of vibrational heterogeneity across the sample \cite{agarwal2012spatially}. This is particularly useful for studying biomaterials, polymers, and composite structures, where vibrational properties vary locally due to structural disorder or hierarchical organization.

For materials with inherently weak Raman scattering or requiring enhanced surface sensitivity, surface-enhanced Raman spectroscopy (SERS) exploits localized surface plasmon resonances in metallic nanostructures to dramatically increase Raman cross-sections. This technique enhances the detection of molecular interactions at interfaces and enables the identification of low-concentration species in heterogeneous systems \cite{cialla2012surface}. Similarly, tip-enhanced Raman spectroscopy (TERS) combines atomic force microscopy (AFM) with Raman scattering to achieve sub-nanometer spatial resolution, overcoming the issue of the diffraction limit and allowing for the study of vibrational properties at individual domain boundaries and defects \cite{verma2017tip}.

Extending Raman spectroscopy into the low spatial frequency range ($\sim$1–50 cm$^{-1}$) allows for the detection of collective excitations, lattice vibrations, and soft phonon modes that can provide insights into interlayer interactions, mechanical properties, and phonon transport in complex materials \cite{ji2016low}. Additionally, temperature-dependent and time-resolved Raman spectroscopy enables to probe dynamic disorder and anharmonic effects by tracking vibrational changes under controlled thermal conditions or ultrafast laser excitation \cite{lucazeau2003effect}.

Given the inherent complexity of vibrational interactions in hierarchical systems, computational modeling plays a crucial role in spectral interpretation \cite{romaniuk2025influence}. Density functional theory (DFT) simulations and machine-learning-based spectral analysis help identify Raman-active modes, predict vibrational spectra, and disentangle overlapping spectral features from different structural components.

\subsubsection{Nuclear Magnetic Resonance (NMR)} spectroscopy exploits the interaction between nuclear spins and an external magnetic field to probe local atomic environments and molecular dynamics \cite{gerothanassis2002nuclear}. Certain atomic nuclei, such as $^1$H, $^{13}$C, $^{29}$Si possess nonzero spin and behave like microscopic magnetic dipoles. When placed in a strong static magnetic field $B_0$, these nuclei align with or against the field, creating discrete energy levels. A radiofrequency (RF) pulse applied perpendicular to $B_0$ excites the nuclei to higher energy states. As they relax back to equilibrium, they emit RF signals, which are then detected and analyzed. The resonance condition follows the Larmor equation:

\begin{eqnarray}
\omega_0 = \gamma B_0,
\label{nmr}
\end{eqnarray}

where $\omega_0$ is the resonance frequency and $\gamma$ is the gyromagnetic ratio (a nucleus-dependent constant). The chemical environment around the nuclei produces chemical shifts in their resonance frequency, while interactions such as dipole-dipole coupling, J-coupling, and quadrupolar effects provide insight into molecular structure, dynamics, and intermolecular interactions.

In crystalline solids, solid-state NMR is widely employed for atomic-resolution structural characterization. It can distinguish between different polymorphs, crystallographic sites, and chemical compositions based on chemical shifts, while relaxation times (T$_1$, T$_2$) reveal different kinds of molecular motions \cite{li2021solid}. The technique is also used in quantum materials to probe local electronic environments, charge ordering, and magnetic excitations \cite{sugiyama1998antiferromagnetic,sugiyama1993nmr}. To improve spectral resolution, solid-state NMR often employs magic-angle spinning (MAS), which works by rapidly rotating the sample at 54.74$^{\circ}$ to the magnetic field. This allows to average out anisotropic interactions like dipolar coupling and chemical shift anisotropy, thereby sharpening spectral lines \cite{polenova2015magic}.

However, when applied to hierarchical and disordered systems, NMR also faces several challenges. The absence of long-range order leads to broad and overlapping spectral features, making it difficult to assign distinct chemical environments. Structural heterogeneity results in a wide distribution of local interactions, while slow molecular dynamics and multiscale interactions obscure clear spectral separation. Therefore, NMR signals here often reflect a mix of multiple relaxation processes occurring on different timescales.

To overcome these limitations, advanced pulse sequences such as dipolar recoupling \cite{schnell2004dipolar}, double-quantum NMR \cite{gregory1997determination}, and heteronuclear correlation (HETCOR) \cite{vasavi2011heteronuclear} enhance resolution and selectively probe different local environments. Relaxation and diffusion NMR methods extract molecular mobility and distinguish between rigid and flexible domains \cite{conibear2014insights}, while multi-dimensional correlation spectroscopy (e.g., HSQC, NOESY, CP-MAS) helps disentangle complex spectral overlaps in amorphous systems \cite{spiess1997multidimensional}. Additionally, dynamic nuclear polarization (DNP) significantly enhances signal intensity in materials with weakly interacting nuclei, improving sensitivity for detecting rare structural motifs \cite{abhyankar2021challenges}. Finally, low-field and portable NMR systems enable in-situ characterization of structure and dynamics \cite{grootveld2019progress}.

\subsubsection{Muon Spin Rotation ($\mu$SR)} is a powerful spectroscopic technique that probes local magnetic fields within materials by exploiting the properties of muons: elementary particles with spin $\frac{1}{2}$, particularly sensitive to local magnetic environments \cite{blundell2021muon}. In a typical $\mu$SR experiment, spin-polarized positive muons are implanted into a sample, where they interact with internal magnetic fields before decaying into positrons. The emission direction of these positrons is correlated with the muon's spin orientation at the time of decay, providing information about the internal field dynamics through the time evolution of the muon spin polarization.

Traditionally, $\mu$SR has been widely applied to investigate quantum materials, superconductors, and magnetic systems. In superconductors, it provides insight into flux-line lattices, penetration depths, and unconventional pairing mechanisms \cite{blundell2004muon,nocerino2022superconducting,grinenko2021split,di2015intrinsic}, while in magnetic materials, it detects spin fluctuations, long-range order, and magnetic phase transitions \cite{nocerino2024cr,sugiyama2020magnetic,papadopoulos2022influence}. The technique excels in measuring extremely weak magnetic fields and distinguishing between static and dynamic local field variations, making it an invaluable tool for studying emergent magnetic and superconducting phenomena. Beyond magnetism and superconductivity, $\mu$SR is also a highly effective probe for studying free radical formation and reactivity, particularly through Muon Spin Relaxation and Avoided Level Crossing (ALC) spectroscopy \cite{mckenzie2013positive}, as well as ion-diffusion mechanisms in energy materials \cite{sugiyama2009li,nocerino2024ion,palm2021ion,benedek2020quantifying}. When implanted into molecular systems, positive muons can form muonium ($\mu^+ e^-$), an exotic light isotope of hydrogen. Due to its high reactivity, muonium readily adds to unsaturated molecular bonds, creating muoniated radicals, which behave similarly to hydrogenated radicals. By analyzing the muon's hyperfine interactions and relaxation dynamics, $\mu$SR provides insights into radical structures, reaction kinetics, and spin interactions, making it a valuable tool for studying radiation chemistry, oxidative stress mechanisms, and organic photochemistry \cite{hubbard2004avoided,heming1989separation,wang2017temporal}. Furthermore, depth-resolved $\mu$SR using energy-selected muons, enables the study of interfacial effects and magnetic heterogeneities across layered structures and thin films \cite{brahlek2023emergent,pratt2016nanoscale}.

While extensively applied to study magnetic properties in various systems, the application of $\mu$SR to hierarchical materials is less common. Here, the lack of translational symmetry in amorphous and multiscale structures leads to spatially inhomogeneous magnetic environments, that cause muons to experience a distribution of local fields rather than a well-defined precession frequency. The resulting broadened relaxation spectra and complex spin dynamics cannot easily be disetangled using conventional muon-site calculations \cite{gingras1997muon}.

To address these challenges, $\mu$SR experiments in hierarchical materials could benefit from complementary analysis methods and tailored experimental setups. In particular, advanced computational modeling, including stochastic simulations and density matrix approaches, would be needed to disentangle overlapping relaxation contributions and extracting meaningful information about local field distributions. 

\subsubsection{Inelastic neutron scattering (INS) and inelastic X-ray scattering (IXS)} are powerful tools for probing elementary excitations in matter, providing insight into phonon, magnon, and electronic interactions by measuring energy and momentum transfer between the incident radiation and the sample \cite{squires1996introduction,schulke2007electron}. Both methods rely on the fundamental principle that when neutrons or X-rays scatter inelastically from a material, they exchange energy $\hbar \omega$ and momentum $\hbar \boldsymbol{Q}$ with collective excitations, enabling the reconstruction of dispersion relations and dynamic correlations. The quantity measured, that is, the dynamic structure factor $S(\boldsymbol{Q}, \omega)$, reflects how excitations propagate and dissipate energy within a system.

Traditionally, these techniques have been instrumental in directly mapping well-defined phonon dispersion curves in crystalline solids, where lattice periodicity ensures the existence of sharp vibrational and electronic modes. Inelastic neutron scattering is particularly suited for studying phonons due to the strong interaction of neutrons with atomic nuclei, and their ability to probe bulk properties without significant absorption, making it ideal for light-element materials and magnetism-related studies \cite{halloran2025connection}. IXS, on the other hand, benefits from the high brilliance of synchrotron X-ray sources and is especially effective in systems where neutron experiments are limited by sample size or weak neutron scattering cross-sections. Additionally, resonant inelastic X-ray scattering (RIXS) is a specialized variant that enhances sensitivity to specific electronic excitations, particularly in transition metal oxides and strongly correlated electron systems \cite{kotani2001resonant}.

However, when applied to hierarchical and disordered materials, both INS and IXS can face significant challenges due to the lack of long-range periodicity. Instead of sharp phonon peaks, vibrational spectra broaden, and the notion of a well-defined wavevector $\boldsymbol{Q}$ loses meaning as coherence lengths become comparable to, or smaller than, the probed wavelengths. Additionally, hierarchical structures introduce multiple scattering pathways and hybridizing vibrational modes across different scales, making it difficult to isolate specific excitation contributions. This complexity is further exacerbated by the occurrence of diffusive dynamics and anharmonic interactions (induced by, e.g., the presence of moisture) which typically contribute to complex energy dissipation mechanisms in soft matter, biomaterials, and porous media.

To overcome these challenges, these measurements can be adapted to the requirements of the specific experiments and combined with other methods (e.g., SAS \cite{aahl2024moisture}). The use of isotope substitution (D for H) and polarization analysis is also widely exploited to enhance sensitivity to specific atomic interactions, enabling targeted investigations of complex materials \cite{berrod2018inelastic}. In IXS, high-resolution spectrometers allow measurements of low-energy excitations to connect vibrational dynamics with slower relaxation processes \cite{baron2020high}. RIXS, by exploiting element-specific resonances \cite{mukkattukavil2022resonant}, could provide a means of selectively probing local electronic and magnetic excitations even in structurally complex materials. Moreover, complementary real-space techniques, including neutron/X-ray pair distribution function analysis, aid in interpreting excitation spectra in disordered and hierarchical systems by providing insight into short-range structural correlations. Computational modeling, particularly molecular dynamics simulations and first-principles calculations, are needed in this context to be able to extract meaningful physical insights from broadened spectra and disentangle contributions from different hierarchical levels \cite{karnbach2021molecular}.


\subsubsection{Pump-probe spectroscopy} is an ultrafast method used to study the transient dynamics of electronic excitations in materials by tracking their evolution on femtosecond to nanosecond timescales \cite{khitrova1988theory}. It operates by using two ultra short optical pulses generated by a laser: a pump pulse excites the system, creating an out-of-equilibrium state, while a probe pulse, delayed by a controlled time interval, measures the system's response. By varying the time delay between the pump and probe, it is possible to reconstruct the relaxation pathways and interactions governing the material’s dynamic response. The technique is particularly powerful in investigating electron-phonon coupling, non-equilibrium phase transitions, carrier dynamics in semiconductors, and ultrafast spin or lattice interactions \cite{wu2024ultrafast,}.

Indeed, pump-probe spectroscopy has been traditionally applied in crystalline materials where well-defined band structures and phonon dispersions facilitate a clear interpretation of energy transfer mechanisms. In semiconductors and correlated electron systems, time-resolved reflectivity or transmission measurements provide direct insights into electronic relaxation, recombination processes, and collective excitations \cite{kaindl2017time}. In metals and superconductors, pump-probe experiments resolve quasiparticle lifetimes and phonon dynamics, revealing critical information about electron-lattice interactions \cite{lobo2005photoinduced}. Angle-resolved photoemission spectroscopy (TR-ARPES), an extension of this technique, provides direct momentum-resolved insights into ultrafast carrier dynamics by mapping transient electronic band structures \cite{boschini2024time}.

However, in hierarchical and disordered materials, the absence of periodicity complicates the interpretation of ultrafast measurements. Unlike in crystalline systems, where relaxation pathways are well-defined and dispersion relations can accurately predict energy propagation, hierarchical materials possess heterogeneous energy landscapes where charge, vibrational, and magnetic excitations couple across multiple scales. This leads to spatially dependent relaxation mechanisms, broad distributions of recombination timescales, and non-uniform energy dissipation. In amorphous materials, diffusive rather than ballistic transport dominates, making it difficult to extract meaningful transport parameters from standard pump-probe measurements. Additionally, in soft matter, biomaterials, and hybrid materials, ultrafast dynamics involve coupled degrees of freedom, such as electronic, vibrational, and molecular conformational changes that evolve on distinct timescales, complicating data interpretation. Nonetheless, it's important to note that pump-probe spectroscopy has been effectively applied to study soft complex systems such as conjugated polymers \cite{kee2014femtosecond}.

To address the challenges posed by hierarchical materials, adaptations of pump-probe spectroscopy may include spatially resolved ultrafast microscopy, which allows for localized probing of relaxation dynamics in structurally heterogeneous materials \cite{wen2019spatially}. Multi-probe techniques incorporating complementary measurements, such as time-resolved X-ray or electron diffraction, help disentangle the coupled dynamics of electrons, phonons, and structural rearrangements in disordered systems . Furthermore, polarization- and wavelength-dependent pump-probe spectroscopy enables selective excitation of specific vibrational or electronic states, reducing spectral congestion from overlapping signals. In materials where transport properties vary across length scales, integrating pump-probe experiments with real-space imaging techniques, such as near-field optical spectroscopy, can provide spatially resolved insights into ultrafast energy transport \cite{grumstrup2015pump}.

\subsection{Free Electron Laser (FEL)} is an advanced light source that generates extremely intense and coherent radiation spanning a wide range of wavelengths, from infrared to X-rays \cite{freund1992principles}. Unlike conventional lasers, which rely on electronic transitions in atoms or molecules, FELs operate by accelerating a high-energy electron beam through a series of alternating magnetic fields, known as undulator. As the relativistic electrons traverse this periodic magnetic structure, they undergo oscillatory motion and emit synchrotron radiation. Through a self-amplified spontaneous emission (SASE) process, the emitted radiation and the electron beam interact, leading to coherent amplification and the production of ultrashort, high-brilliance pulses with femtosecond to attosecond temporal resolution.

Traditionally, FELs have been instrumental in studying ultrafast electronic, vibrational, and structural dynamics in ordered materials. X-ray free electron lasers (XFELs) provide femtosecond-scale snapshots of transient states in solid-state systems, enabling direct visualization of phase transitions, electron-phonon interactions, and bond breaking or formation \cite{lee2023observing}. In crystallography, FELs can enable single-shot diffraction experiments, revealing the structure of biological macromolecules, but also quantum materials and complex solids, with minimal radiation damage- \cite{huang2020xfel}. 

Nonetheless, when applied to disordered materials FEL experiments face unique challenges arising from the intrinsic lack of long-range order and the multiscale nature of interactions. In periodic crystals, FEL-based diffraction and scattering techniques rely on well-defined Bragg reflections and reciprocal-space representations, whereas in amorphous or hierarchical systems, diffuse scattering dominates, making it difficult to extract clear structural information. To overcome these limitations, spatially and temporally selective FEL techniques have been developed. Coherent diffraction imaging (CDI) \cite{miao2011coherent} and ptychography \cite{kharitonov2022single} bypass the need for periodic order by reconstructing real-space electron density maps directly from the scattered X-ray intensity, making them particularly suitable for disordered materials. Ultrafast X-ray photon correlation spectroscopy (XPCS) \cite{roseker2018towards} exploits the coherence of FELs to probe slow dynamics and mesoscale fluctuations in disordered systems.

Despite these advancements, challenges persist. In XFEL experiments the use of small samples is essential due to the intense and focused nature of the X-ray pulses. This is because the microscopic size of the XFEL beam necessitates precise targeting of the sample, which is more effectively achieved with smaller sample volumes \cite{amann2023liquid}. This implies that hierarchical structures, that span several lengthscales in size, are not suitable for this kind of measurement. Moreover, the highly focused pulses can cause significant radiation damage, leading to the destruction of the exposed sample area. Indeed, experiments often have to employ serial femtosecond crystallography (SFX), where numerous microcrystals are sequentially introduced into the X-ray beam. This approach allows each crystal to be exposed only once before being replaced by a new one, so that high-quality diffraction data can be collected before cumulative radiation damage affects the samples (adhering to the "diffraction-before-destruction" principle) \cite{lehmkuhler2018dynamics,martiel2019strategies}.

\subsubsection{Terahertz (THz) spectroscopy} operates in the electromagnetic spectrum between infrared and microwave frequencies (0.1–10 THz), probing low-energy excitations such as phonons, molecular vibrations, charge transport, and polaritonic interactions \cite{beard2002terahertz}. In a typical THz time-domain spectroscopy (THz-TDS) setup, an ultrashort laser pulse excites a photoconductive emitter, generating a broadband THz pulse that interacts with the sample. The transmitted or reflected signal is then detected coherently, yielding phase-resolved information on absorption and refractive index variations. Because THz photons correspond to meV-scale energies, this technique is particularly useful, and widely employed in solid-state physics, for studying long-wavelength lattice vibrations, hydrogen bonding networks, charge dynamics in semiconductors, and collective excitations in correlated electron systems \cite{kimura2012infrared}. In periodic lattices, THz absorption spectra exhibit well-defined peaks corresponding to distinct interband transitions and phonon modes \cite{kuehn2011two,kim2023terahertz}. Additionally, THz spectroscopy is particularly well-suited for well-ordered soft matter and biological systems, where it provides sensitivity to intermolecular interactions, hydration dynamics, and low-energy excitations associated with protein folding \cite{mancini2022terahertz}, membrane fluctuations, and charge transport in organic materials \cite{giannini2022charge}.

In hierarchical and disordered systems, however, the lack of long-range order and the multiscale nature of vibrational interactions lead to broad spectral distributions, disorder-induced shifts in vibrational energies, and the overlap of coherent lattice vibrations with diffusive or relaxational processes, making data interpretation more complex.

To address these challenges, spatially and temporally resolved THz techniques have been developed. Time-resolved THz spectroscopy captures ultrafast carrier and vibrational dynamics in disordered materials, revealing non-equilibrium transport mechanisms that may be obscured in static measurements \cite{wang2016time}. Near-field THz imaging enhances spatial resolution beyond the diffraction limit, enabling the investigation of local vibrational heterogeneities in, e.g., metamaterials \cite{hale2023near}. Furthermore, computational modeling of disordered phonon spectra aids in the interpretation of broad spectral features, providing insight into local vibrational density of states and disorder-induced mode coupling \cite{ruggiero2020invited}.

\subsubsection{Brillouin spectroscopy} is a powerful tool for probing acoustic phonons, viscoelastic properties, and mechanical moduli by measuring the inelastic scattering of light from thermally excited phonons \cite{meng2016seeing}. In a typical Brillouin scattering experiment, a monochromatic laser beam interacts with a material, and the scattered light undergoes a frequency shift due to the Doppler effect associated with acoustic phonon propagation. The frequency shift ($\Delta \nu$) is directly related to the sound velocity $v$ and the elastic moduli of the material through the equation:

\begin{eqnarray}
\Delta \nu = \frac{2nv}{\lambda} \sin{\theta/2},
\label{brillouin}
\end{eqnarray}

where $n$ is the refractive index of the material, $\lambda$ is the laser wavelength, and $\theta$ is the scattering angle. By analyzing these frequency shifts, Brillouin spectroscopy provides non-contact measurements of mechanical properties, phonon dispersion, and viscoelastic behavior in a wide range of materials.

Brillouin spectroscopy has been used in solid-state physics to study acoustic phonon dispersions in single crystals and polycrystalline materials \cite{kang2021elastic,reed2022acoustic}, as well as in soft matter to investigate rheological properties, viscoelasticity, and phase transitions \cite{rodriguez2023network,kharmyssov2024brillouin}. In ordered solids, the technique provides direct access to elastic tensor components, revealing anisotropic mechanical properties and phonon lifetimes. However, in hierarchical and disordered systems, Brillouin spectroscopy encounters significant challenges due to mode hybridization and localization, as well as strong scattering losses. The complex structural organization of these kinds of systems often leads to a broadening of Brillouin peaks, making it difficult to extract meaningful phonon dispersion relations. Additionally, in highly porous or heterogeneous materials, multiple scattering and confinement effects obscure the intrinsic acoustic properties of the system \cite{bottani2018brillouin}.

To address these challenges, angle-resolved and high-resolution Brillouin spectroscopy techniques have been developed to separate overlapping modes and extract detailed information on phonon lifetimes and damping mechanisms \cite{mattarelli2020relevant}. Additionally, Brillouin light scattering (BLS) microscopy enables spatially resolved mechanical characterization, which allows to map local variations in stiffness and viscoelastic properties across complex structures \cite{singaraju2019brillouin}. In hierarchical materials, where distinct length scales contribute to mechanical behavior, multi-modal approaches combining Brillouin spectroscopy with Raman, X-ray, or neutron scattering provide complementary insights into phonon transport and structural dynamics \cite{alunni2021brillouin}.

\subsubsection{Quasi-Elastic Neutron Scattering (QENS)} is a powerful technique for investigating dynamical processes, in particular diffusive motions, molecular reorientations, and relaxational phenomena \cite{telling2020practical}. The fundamental principle of QENS is based on the inelastic scattering of neutrons, where a neutron transfers a small amount of energy $\hbar \omega$ to a system without exciting well-defined phonon modes. Instead of sharp energy exchanges, characteristic of coherent inelastic neutron scattering (INS), QENS spectra exhibit broadening around the elastic peak, reflecting a distribution of low-energy excitations associated with stochastic or thermally activated processes. Mathematically, the observed scattering intensity is, also in this case, described by the dynamic structure factor $S(\boldsymbol{Q}, \omega)$ which provides insight into both the spatial ($\boldsymbol{Q}$) and temporal ($\omega$) scales of molecular motions. The width of the quasi-elastic signal $\gamma$ relates directly to characteristic diffusion coefficients or relaxation times through models such as Fickian diffusion or jump processes.

In solid-state systems, QENS has traditionally been used to study atomic and molecular diffusion, proton transport, and relaxation processes in crystalline hydrates, intercalation compounds, and energy materials such as battery electrolytes \cite{lyonnard2010perfluorinated,desmedt2011dynamics,juranyi2015dynamics,triolo2001dynamic}. The technique is particularly valuable for probing hydrogen dynamics, as neutrons are highly sensitive to hydrogen nuclei due to their large scattering cross-section. In single crystals and ordered materials, QENS spectra often exhibit well-defined diffusion mechanisms, with characteristic relaxation times that can be described by established transport models.

However, when applied to hierarchical and disordered systems, QENS signals exhibit broad and overlapping spectral features due to the coexistence of multiple dynamic regimes. In hierarchical materials and soft matter, molecular motions also occur across multiple timescales, ranging from picoseconds for local librations to nanoseconds or longer for collective relaxations. This simultaneous contribution of multiple dynamic processes leads to complex QENS spectra, where distinguishing between localized vibrational motions, rotational relaxation, and long-range diffusive transport becomes nontrivial.

To overcome these challenges, QENS experiments in hierarchical and disordered systems rely on energy-resolved neutron spectrometers with variable resolution, which allow to separate fast, localized motions from slower, long-range transport mechanisms \cite{crupi2003structure,liu2006quasielastic}. Additionally, the use of temperature-dependent QENS enables the characterization of activation barriers and dynamic crossovers \cite{diallo2015translational}. Complementary real-space modeling approaches, including molecular dynamics simulations and jump-diffusion models, assist in interpreting QENS spectra \cite{yasuda2019molecular}. Furthermore, QENS studies of confined systems, such as water in nanoporous materials or polymer electrolytes \cite{lyonnard2012neutrons}, employ contrast variation techniques using deuteration, selectively highlighting specific molecular species to isolate their contribution to the overall dynamics \cite{kruteva2021dynamics}.

\section{Connecting Scales and the Role of Machine Learning in Overcoming the Multiscale Characterization Challenge}\label{ml}

As seen above, no single experimental technique can capture the full range of interactions of interest spanning multiple length/time scales and, therefore, a multimodal approach combining together several techniques, is currently the preferred route for multiscale characterization of complex structures. However, in this way, the connections between structural elements and interactions at different scales cannot truly be directly observed, and instead, they have to be inferred. Therefore, understanding emergent behavior in structurally complex or hierarchically organized materials remains a central challenge in modern materials science. 

In recent years, new characterization techniques have been specifically designed and adapted to provide structural information across multiple length scales in a single measurement, addressing the limitations of traditional methods that focus on either local atomic-scale details or macroscopic properties. Techniques such as small-angle X-ray scattering tensor tomography (SAXS-TT) \cite{liebi2015nanostructure}, dark-field X-ray and neutron imaging \cite{pfeiffer2008hard,strobl2008neutron}, scanning electron diffraction \cite{nero2024nanoscale}, and X-ray linear dichroic tomography \cite{apseros2024x} have emerged as powerful tools for mapping structural features from the nanoscale to the meso- and macro-scale in hierarchical materials. These methods, by combining scattering and imaging principles, allow to capture structural organization across hierarchical levels within a single measurement.

SAXS-TT extends the principles of SAXS by acquiring angular-dependent scattering information while rotating the sample, allowing for three-dimensional reconstruction of nanostructural anisotropy. Here, a sample is scanned through a X-ray beam, and 2D scattering patterns are collected at each scanning point. This scanning process is repeated across various orientations and tilting of the sample with respect to its tomographic axis, and a 3D reciprocal space map is reconstructed combining the 2D patterns for each element of volume in the beam path. In this way SAXS-TT reconstructs the full orientation distribution function (ODF) of nanostructures in volumes on the order of a few cubic millimeters, with a spatial resolution of 10–50 $\mu$m and a structural resolution down to a few nanometers \cite{georgiadis2021nanostructure,murer2021quantifying}. 

Dark-field X-ray and neutron imaging utilize ultra-small-angle scattering (USAXS) to extract structural contrast beyond the pixel size of conventional radiography. These methods operate in a scattering vector range of 10$^{-3}$ to 10$^{-2}$ nm$^{-1}$, corresponding to sensitivity for structures ranging from 10 nm to several micrometers, far beyond the diffraction limit of direct imaging techniques. By inserting a structured grating or a crystal analyzer into the beam path, the method measures the differential scattering power of a sample, enabling spatially resolved imaging of sub-resolution microstructures in sample volumes up to several cubic centimeters \cite{viermetz2022dark,valsecchi2020characterization,busi20243d}.

At even higher spatial resolutions, scanning electron diffraction (SED) combines the high spatial resolution of scanning electron microscopy (SEM) with the crystallographic sensitivity of electron diffraction. In SED, a focused electron beam (typically 1–10 nm in diameter) is raster-scanned across the sample while acquiring a diffraction pattern at each point, generating a "4D" dataset (two real-space and two reciprocal-space dimensions). This allows to map local crystal orientations, strain fields (via shifts in diffraction spot positions), and phase distributions with a spatial resolutions of 1 to 5 nm with angular resolutions of 0.1–1 mrad over sample areas spanning hundreds of microns. It is particularly advantageous for polycrystalline and nanostructured materials, where conventional diffraction methods struggle to resolve grain boundary interactions and lattice distortions over extended sample regions \cite{tovey2020scanning}.

X-ray linear dichroic tomography (XLDT), on the other hand, exploits the anisotropic absorption of linearly polarized X-rays to reconstruct 3D maps of electronic and chemical anisotropies with spatial resolutions of 50–500 nm and energy resolutions of 0.1 eV. By systematically varying the polarization angle and sample orientation, XLDT measures the differential absorption of X-rays aligned parallel and perpendicular to specific molecular or electronic structures. This method quantitatively resolves tensorial absorption coefficients in the sample, revealing nanoscale orientation distributions of functional groups, charge density anisotropies, and magnetic domain structures. XLDT operates with sub-micron to micron spatial resolution, making it a powerful tool for studying oriented molecular systems (e.g., liquid crystals), interfacial chemistry in energy storage devices, and magnetization textures in functional materials \cite{apseros2024x}.

Beyond these, several emerging techniques are explored for multiscale characterization, accompanied by instrumental adaptations that enable measurements under environmental conditions not typically addressed by such methods, for example, moisture-dependent studies \cite{nocerino2025design}. Coherent diffraction imaging (CDI) \cite{clark2012high} and ptychography \cite{pfeiffer2018x} reconstruct real-space structural information from diffraction patterns, providing lensless, high-resolution imaging of strain fields and nanoscale heterogeneities. X-ray photon correlation spectroscopy (XPCS) exploits wave coherence to probe slow dynamical fluctuations over extended correlation lengths, allowing to detect mesoscale transport and relaxation phenomena in disordered systems \cite{perakis2020towards}. Bragg coherent diffraction imaging (BCDI) enables 3D mapping of internal strain and defect distributions \cite{sun2024bragg}.

For systems with complex microstructures, multi-scale computed tomography (CT) combines different X-ray imaging modes to zoom into regions of interest, preserving both large-area context and fine structural details \cite{pini2016moving}. Similarly, 4D-scanning transmission electron microscopy (4D-STEM) captures diffraction data at each scanned pixel, providing an unprecedented view of strain gradients, phase distributions, and electronic structure variations within nanostructured materials \cite{guo20234d}. Neutron spin echo (NSE) spectroscopy is a powerful technique that resolves ultra-slow molecular dynamics by measuring the intermediate scattering function, $F$($Q$,$t$), over time scales ranging from picoseconds to hundreds of nanoseconds. This capability enables the investigation of slow relaxation processes and diffusive motions in materials such as glasses, polymer melts, complex fluids, and microemulsions \cite{richter2005neutron}.

The large and complex datasets generated by these multiscale techniques present new challenges in data interpretation, which advanced computational modeling and machine learning approaches can help address by identifying meaningful correlations between microscopic structures and macroscopic physical properties.


Traditional multiscale modeling approaches rely on either bottom-up or top-down methodologies, where information is transferred across scales through predefined constitutive relations or homogenization schemes. However, these methods face inherent limitations: bottom-up approaches (e.g., molecular dynamics feeding into continuum models) are computationally prohibitive for large systems, while top-down methods often oversimplify fine-scale interactions \cite{karabasov2014multiscale}. 
Machine learning (ML) has transformed multiscale modeling with an approach that involves learning complex structure-property relationships directly from data, making it possible to establish predictive models without explicitly solving every equation at each scale \cite{alber2019integrating}. This capability is particularly useful in materials science, where multiscale interactions often lead to emergent behavior that is difficult to describe analytically.
Central to ML's success in this domain are artificial neural networks: computational models inspired by the human brain's interconnected neuron structures. In fact, the formulation of mathematical models that underpin modern neural networks began with the idea to address a very ambitious researh question: understand and replicate the human brain's computational capabilities. In 1943, neurophysiologist Warren McCulloch and mathematician Walter Pitts introduced a pioneering model that conceptualized neurons as binary threshold units. They demonstrated that networks of such simplistic neurons could, in principle, perform computations equivalent to those of a Turing machine, laying the groundwork for the computational theory of mind \cite{neurons}. The McCulloch-Pitts model abstracted neural activity into a framework where each neuron received multiple inputs, computed a weighted sum, and produced an output based on a threshold function. With this formulation it is easy to see that, by emulating the hierarchical organization observed in natural systems, interconnected simple units in a network structure can perform complex computations to the point that information processing, similar to the one of the human brain, is achievable artificially.

Building upon this foundation, in 1958 Frank Rosenblatt developed the perceptron, aiming to create a machine that could learn from experience \cite{block1962perceptron}. The perceptron introduced adjustable weights, allowing the model to learn decision boundaries from input data. Mathematically, the perceptron's operations relied on linear algebra: input vectors were multiplied by weight matrices, and the resulting vector was transformed via an activation function to produce outputs. This linear combination of inputs and weights formed the basis for pattern recognition tasks. Therefore, in practice, perceptron computed a weighted sum of inputs and passed the result through a step activation function to produce an output. Here, the machine was able to learn by adjusting the weights based on the error between the predicted and actual outputs, as quantified by the loss function. This function measures the discrepancy between the model's predictions and the true values, guiding the weight updates to minimize errors. Such a mechanism is foundational to modern learning algorithms.

However, despite its promise, perceptron was not able to solve problems that were not linearly separable (e.g., the XOR problem), among other limitations. This challenge highlighted the need for more complex architectures and learning algorithms, which led to the resurgence of neural network research in the 1980s, with the development of the backpropagation algorithm \cite{werbos2005applications,rumelhart1986learning}. This algorithm enabled the training of multi-layer perceptrons by efficiently computing gradients needed for weight updates across multiple layers. Backpropagation utilizes calculus to compute the gradient of the loss function with respect to each weight in the network. By applying the chain rule for composite functions, it propagates errors backward from the output layer to the input layer, allowing for the adjustment of weights in a manner that minimizes the overall error. This process iteratively updates weights using gradient descent, which is a method rooted in optimization theory \cite{haji2021comparison}. However, neural networks at this point would assign fixed values to their weights, leading to point estimates in their outputs that could not adequately capture the variability of real-world data.

The integration of probability and statistics into neural network models enhanced further their capability to handle uncertainty and make probabilistic predictions \cite{sarle1994neural}. By modeling weights as probability distributions rather than fixed values, these networks could express confidence levels in the likelihood of their outputs. This probabilistic approach advanced backpropagation by incorporating prior knowledge and quantifying uncertainty, leading to more robust performance in tasks like classification and regression, where understanding the underlying data distribution is a necessary step \cite{goan2020bayesian}.

Grounded in these mathematical foundations, the field has progressed from early neural networks used for basic pattern recognition to sophisticated architectures capable of learning high-dimensional mappings between physical scales. The first wave of ML applications in physics and materials science relied on fully connected neural networks (FCNNs) for regression and classification tasks \cite{ganju2018property}. These networks, composed of stacked layers of artificial neurons, were trained to approximate material properties based on predefined input features. Early implementations helped automate structure-property predictions in simple systems, such as estimating mechanical properties from basic chemical compositions or crystal structures \cite{zheng2018machine}. However, these models struggled with high-dimensional data and lacked the ability to capture spatial correlations, making them unsuitable for problems involving complex microstructures. To address spatial correlations in structured data, convolutional neural networks (CNNs) emerged as a powerful tool \cite{chauhan2018convolutional}. Originally developed for image recognition, CNNs exploit local connectivity and weight sharing to efficiently extract hierarchical features from structured input, such as imaging data, X-ray scattering patterns, and molecular dynamics simulations \cite{bullock2019xnet,liu2019convolutional,chew2020fast}. For this reason, in materials science, CNNs enabled the automatic classification of microstructures, defect detection, and contrast-based segmentation in tomography datasets. However, CNNs assume a fixed grid structure (e.g., pixel grids in imaging data), making them less suited for irregular or non-Euclidean data, such as non-crystalline atomic networks, porous media, or mesoscale architectures. To overcome this limitation, graph neural networks (GNNs) were introduced, which model topological relationships between structural elements rather than relying on a predefined grid \cite{wu2020comprehensive}. GNNs enabled more realistic material representations, capturing atomic connectivity in molecular systems, mechanical stress propagation, and even phase transitions \cite{reiser2022graph,moradzadeh2023topology}. These models provided a more natural way to incorporate topological and relational data, but they still required extensive labeled datasets and struggled with generalization across vastly different length scales.

These data-driven surrogate models, widely used in computational science to approximate expensive numerical solvers, have been increasingly explored in materials physics for applications such as accelerating molecular dynamics simulations or learning effective constitutive relations directly from high-fidelity simulations or experimental data \cite{pollice2021data}. Instead of explicitly solving governing equations at every scale, these models infer reduced-order descriptions of material behavior, which drastically improves computational efficiency. For example, the Deep Material Network (DMN) enables fast homogenization of heterogeneous materials without exhaustive finite element simulations \cite{liu2019deep}. A key innovation in this regard is represented by Physics-Informed Neural Networks (PINNs), which take a hybrid theory- and data-driven approach, incorporating known physical laws directly into the training process \cite{raissi2019physics}. Instead of relying solely on labeled data, PINNs enforce constraints from governing equations, such as Navier-Stokes equations for fluid flow \cite{cai2021physics}, heat diffusion equations for thermal transport \cite{cai2021physics2}, or elasticity equations for mechanical stress distribution \cite{roy2023deep}. This ensures that ML-generated solutions remain physically consistent even when data is sparse or incomplete. In multiscale materials modeling, PINNs can interpolate between different length scales by embedding fundamental conservation laws as differential operators into the training process. For example, in composite materials, PINNs can predict macroscopic mechanical behavior while respecting microstructural heterogeneity, without requiring expensive full-scale simulations \cite{haghighat2022physics}.

However, conventional surrogate models lack the flexibility to extrapolate to new unknown conditions. This is because they are trained for specific parameter ranges or material compositions, meaning that if the initial conditions or boundary conditions of the system vary, the neural network needs to be re-trained. A new training necessitates building a reasonably large training set for the neural network to provide reliable outputs, which involves computationally expensive simulations.

This problem led to a major breakthrough in the field with the development of neural operators, a new class of ML models that learn continuous mappings between function spaces, enabling predictions for input configurations within the range of the training data, with some extrapolation capabilities \cite{azizzadenesheli2024neural}. These approaches accelerate multiscale simulations by constructing surrogate models that approximate fine-scale interactions without explicitly solving all governing equations at each scale. This eliminates the need for costly iterative computations while maintaining predictive accuracy. Neural operators, such as Deep Operator Networks (DeepONets) \cite{lu2021learning} and Fourier Neural Operators (FNOs) \cite{li2020fourier} to mention two important examples, are designed to learn mappings between entire function spaces rather than discrete data points, enabling them to capture complex multiscale behaviors with significantly reduced computational cost. Compared to conventional numerical solvers, such as finite element and molecular dynamics simulations, which scale as $\mathcal{O}(N^3)$, neural operators can reduce scaling to nearly $\mathcal{O}(N)$ in some cases \cite{alkin2025universal}, by learning continuous representations that bypass expensive computational steps.


DeepONet is a neural network architecture specifically designed to learn nonlinear mappings between input and output functions, making it particularly well-suited for arching over different scales in hierarchical systems. Therefore, in the context of materials physics, DeepONet can be employed to model the effective macroscopic response of a material based on its microstructural features. While traditional neural networks map discrete inputs to discrete outputs, DeepONets encode the entire input function through a trunk network and a branch network. The trunk network processes spatial coordinates (e.g., positions of structural elements in a material), while the branch network encodes external parameters or input conditions (e.g., local stress fields, temperature gradients, or microscopic material properties). The final output is obtained by taking an inner product of the two network representations, such that DeepONet is able to predict the full system response efficiently. This structure allows for fast inference of material behavior across scales, making it applicable to problems such as predicting macroscopic mechanical properties from microstructure \cite{hossain2025virtual,wang2025deeponets} or modeling fluid flow through porous media or non-rigid vessels without explicit resolution of all microstructural details \cite{laudato2025neural,laudato2024high}.

FNOs take a different approach by utilizing fast Fourier transforms (FFT) to represent functions in frequency space, where learning complex spatial relationships becomes more efficient (mutuating the approach of crystallography). This architecture is particularly suited for materials systems where long-range interactions and spatial correlations play a dominant role, such as in transport phenomena or elasticity. Unlike conventional CNNs, which require large receptive fields to capture long-range dependencies, FNOs operate directly in Fourier space, applying neural networks to transformed representations of input data. This method is particularly powerful for modeling systems with strong nonlocal interactions, such as turbulent transport \cite{wang2024prediction}, elastic deformations \cite{li2023fourier}, and wave propagation in heterogeneous media \cite{yang2021seismic}. Here, FNOs is able to significantly reduce memory and computational costs, by learning a compact frequency-space representation, while improving generalization to unseen conditions.

By learning mappings between infinite-dimensional function spaces, neural operator networks have revolutionized the modeling of complex systems and facilitated efficient solutions to problems governed by partial differential equations. The next step at this point would be to enhance the flexibility and accuracy of their outputs. Emerging advancements in this direction are, for instance, Physics-Informed Neural Operators (PINOs) that integrate physical laws directly into the neural architecture (similar to the approach of the PINNs), to make sure that predictions adhere to established principles even in data-scarce scenarios \cite{li2024physics}. Additionally, the incorporation of topological deep learning enables the processing of data on non-Euclidean domains, such as graphs and manifolds, which allows to capture convoluted relationships within complex structures \cite{zia2024topological}. Furthermore, developing latent space neural operators allows for real-time predictions in highly nonlinear and multiscale systems by operating within reduced-dimensional representations \cite{kontolati2024learning}. 
Collectively, these advancements are propelling the field toward more adaptable and precise models, broadening the applicability of neural operators to a diverse range of scientific and engineering challenges.

\section{Conclusions and Perspectives}

Rapid progress in experimental techniques and ML is fundamentally transforming materials science, particularly in the study of structurally complex, hierarchical materials, the most intricate examples of which surround us every day in nature. While traditional methods have provided detailed insights at specific scales, no single technique can fully capture the rich connections that emerge across structural scales. Consequently, numerous fundamental questions remain unexplored, or even unrecognized, and seemingly well-established systems can still hold unexpected surprises when examined more closely.

The integration of multi-resolution experimental methods, such as SAXS-TT, dark-field imaging, electron diffraction, and dichroic tomography, has significantly expanded our ability to probe nanoscale and mesoscale structures within macroscopic samples. However, connecting scales quantitatively remains a fundamental challenge.

Machine learning offers a robust solution to this problem because, beyond dramatically reducing computational costs while retaining predictive accuracy in simulations that would otherwise be prohibitive, ML can also reveal hidden correlations in high-dimensional datasets. This, in turn, opens new avenues for exploring the relationships between structure and function that might otherwise remain unexplored. Moreover, experimental measurements themselves constitute a vast and underutilized source of structured data that can contribute to the training and generalization of ML models, thereby advancing the field in a synergistic way. 

Looking ahead, the integration of ML-based models with multiscale experimental characterization is still in its early stages, but its long-term impact on material science is likely to be profound. As experimental datasets continue to grow and computational power increases, ML-driven multiscale modeling will deepen our fundamental understanding of hierarchical materials. This will ultimately enable the systematic design of novel functional materials with tailored properties.

After all, it seems only fitting that the challenge of understanding hierarchical structures be met with equally hierarchical modeling paradigms.

We are now positioned at the beginning of an era where ML-driven multiscale modeling, guided by advanced experimental methods, bring us closer to a truly predictive framework for materials science; one where scale-connecting models can unify atomic interactions, mesoscale organization, and macroscopic performance, ultimately uncovering the most fundamental and universal principles that govern structure-property relationships in ordinary matter.

\section{Acknowledgments}
The author acknowledges financial support from the SSF-Swedness grant SNP21-0004 and the Foundation Blanceflor 2024 fellow scholarship. The author wishes to thank Dr. Marco Laudato, for his insightful suggestions on section \ref{ml}, and Dr. Giovanni Nocerino, for his valuable support in the composition of Fig. \ref{miller}.

\section*{Data availability statement}
This work does not contain new data.

\textbf{Competing interests} 
The author declares no competing interests.

\section{Bibliography}
 
\bibliography{iopart-num}

\end{document}